\documentclass[hidelinks,journal]{IEEEtran}

\usepackage{cite}
\usepackage{amsmath}

\usepackage{array}

\ifCLASSOPTIONcompsoc
  \usepackage[caption=false,font=normalsize,labelfont=sf,textfont=sf]{subfig}
\else
  \usepackage[caption=false,font=footnotesize]{subfig}

\usepackage{dblfloatfix}
\usepackage{url}

\usepackage{color}
\usepackage{amsthm}
\usepackage{amsmath}
\usepackage{amssymb}
\usepackage{amsfonts}
\usepackage{csquotes}
\usepackage{mathrsfs}
\usepackage{mathtools}

\usepackage{graphicx}
\usepackage{ragged2e}
\usepackage{footnote}
\usepackage[ruled,linesnumbered,algo2e]{algorithm2e}
\usepackage{algorithm}
\usepackage{algpseudocode}
\usepackage[normalem]{ulem}
\usepackage{dirtytalk}
\usepackage{xcolor}

\usepackage{multirow}
\usepackage{enumerate}	
\usepackage{verbatim}
\usepackage{booktabs}
\usepackage{float}
\usepackage{tikz}
\usetikzlibrary{shapes.geometric, arrows}

\usepackage[T1]{fontenc}
\usepackage{authblk}
\usepackage{soul}
\usepackage{comment}

\usepackage[pdfpagelabels,hypertexnames=false,breaklinks=true,bookmarksopen=true,bookmarksopenlevel=2]{hyperref}

\graphicspath{ {./Figures/MotivationAnalysis/} }

\renewcommand{\IEEEQED}

\usepackage{nomencl}
\makenomenclature
\usepackage{pifont}
\begingroup\expandafter\expandafter\expandafter\endgroup
\usepackage{comment}

\usepackage{hyperref}

\makeatletter
\let\original@algocf@latexcaption\algocf@latexcaption
\long\def\algocf@latexcaption#1[#2]{%
	\@ifundefined{NR@gettitle}{%
		\def\@currentlabelname{#2}%
	}{%
		\NR@gettitle{#2}%
	}%
	\original@algocf@latexcaption{#1}[{#2}]%
}
\makeatother

\begin{document}
%
\title{A Hybrid Reactive Routing Protocol for Decentralized UAV Networks}

\author{Shivam~Garg\thanks{S. Garg is with the Computational Science Research Center at San Diego State University (SDSU), USA, shivam\_garg@ieee.org},~\IEEEmembership{Member,~IEEE}, 
\and Alexander~Ihler\thanks{A. Ihler is with the Donald Bren Hall School of Information and Computer Sciences, University of California, Irvine, USA, ihler@ics.uci.edu},~\IEEEmembership{Member,~IEEE}, 
\and Elizabeth~Serena~Bentley\thanks{E. S. Bentley is with the Air Force Research Laboratory, Rome, NY, USA, elizabeth.bentley.3@us.af.mil},~\IEEEmembership{Member,~IEEE}, 
\and Sunil~Kumar\thanks{S. Kumar is with the Electrical and Computer Engineering Department, SDSU, San Diego, USA, skumar@sdsu.edu},~\IEEEmembership{Senior~Member,~IEEE}
\thanks{Distribution A. Approved for public release: Distribution Unlimited: AFRL-2024-4031 on 24 Jul 2024.}
}

\markboth{
}
{S. Garg \MakeLowercase{{et al.}}: A Hybrid Reactive Routing Protocol for Decentralized UAV Networks}
%



\maketitle

\begin{abstract}
Wireless networks consisting of low SWaP (size, weight, and power), fixed-wing UAVs (unmanned aerial vehicles) are used in many applications, such as monitoring, search, and surveillance of inaccessible areas.  
A decentralized and autonomous approach ensures robustness to failures; 
the UAVs explore and sense within the area and forward their information, in a multihop manner, to nearby aerial gateway nodes.
However, the unpredictable nature of the events, relatively high speed of the UAVs and dynamic trajectories cause the network topology to change significantly over time, resulting in frequent route breaks.
A holistic routing approach is needed to support multiple traffic flows in these networks to provide mobility- and congestion-aware, high-quality routes when needed, with low control and computational overhead, using the information collected in a  distributed manner. Existing routing schemes do not address all the mentioned issues.

This paper presents a hybrid reactive routing protocol for decentralized UAV networks called Hyd-AODV. It searches routes on-demand (using a multi-metric route selection), monitors a region around the selected route (the ``pipe''), and proactively switches to an alternative route before the current route's quality degrades below a threshold. 
The impact of pipe width is empirically and theoretically evaluated to find alternate high-quality routes within the pipe and the overhead required to maintain the pipe.
A queue management scheme is also incorporated to prioritize packet transmissions based on their age of information (AoI). Compared to existing reactive routing schemes, the proposed approach achieves higher throughput and reduces the number of route discoveries, overhead, and resulting flow interruptions at different traffic loads, node density, and speeds. Despite having limited network topology information and low overhead and route computation complexity, the proposed scheme achieves a superior throughput to proactive optimized link state routing (OLSR) scheme for different network and traffic settings. The relative performance of reactive and proactive routing schemes is also studied. 
\end{abstract}


\IEEEpeerreviewmaketitle


\section{Introduction}
\label{Introduction}
\begin{figure*}[!th]
	\centering
	\includegraphics[width=0.9\linewidth]{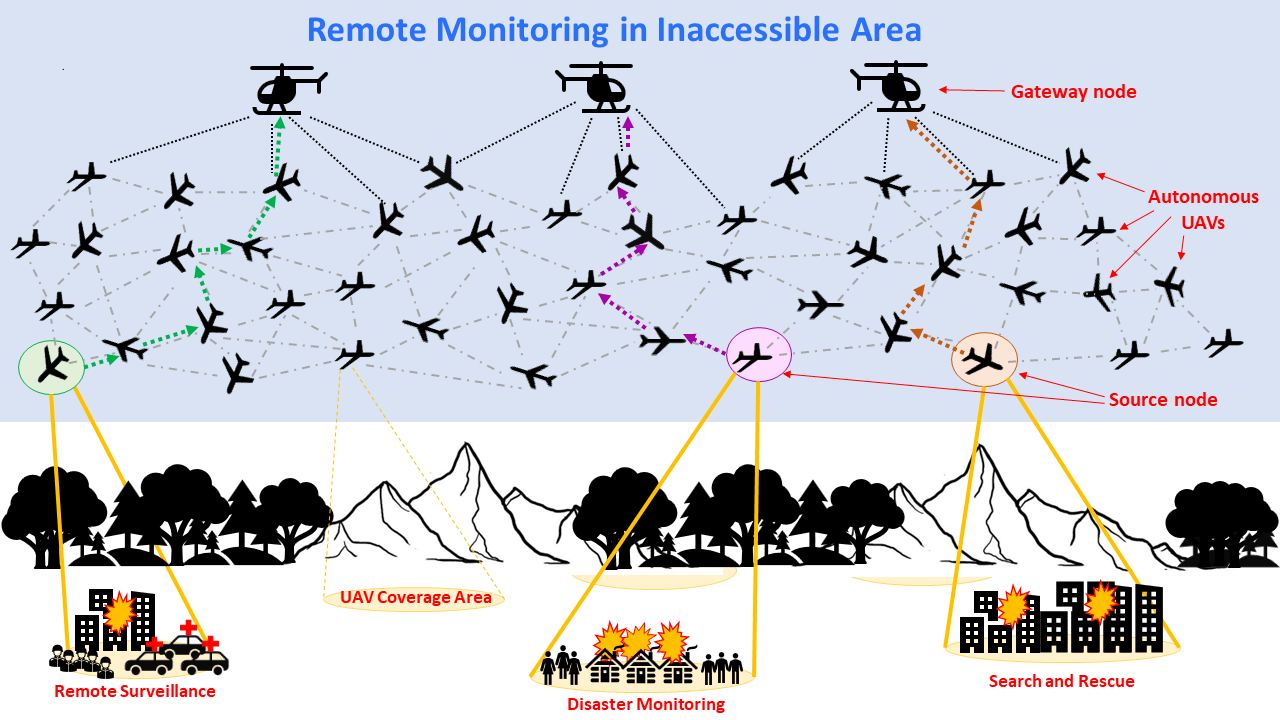}
	\caption {Illustration of autonomous UAV ad-hoc network for remote monitoring of an inaccessible area, where a communication infrastructure is not available.}
	\label{ApplicationScenario}
\end{figure*}

Applications of flying ad-hoc networks (FANETs) include environment sensing, disaster management, surveillance, relay networks, and more \cite{mansoor2023fresh, alam2022survey, rovira2022review, alzahrani2020uav, BAOMDV}. 
Increasingly, unmanned aerial vehicles (UAVs) form a core part of FANETs, thanks to their relatively low cost, rapid deployment, and device autonomy.
A UAV network can easily adapt to dynamic mission requirements, with high reliability and fault tolerance \cite{alam2022survey, rovira2022review}. 
Fig.~\ref{ApplicationScenario} shows an example of remote monitoring in an inaccessible area, where a communication infrastructure is not available.  A decentralized network of low SWaP (size, weight, and power) UAVs, equipped with self-localization and sensing capabilities, can provide reliable sensing and communication for area monitoring and information gathering \cite{alzahrani2020uav,gharib2022lb}. 
The information is time-sensitive and must be transmitted in real time to a remote ground control station (GCS) for coordinating the mission.
However, low SWaP UAVs cannot directly reach the GCS due to their limited signal transmission range.  Therefore, information is forwarded along a multi-hop route to gateway nodes, which can be airborne for longer duration and have a higher transmission range. This enables them to relay the data to the GCS to make informed decisions in real time. 

In an area being monitored, it is difficult to predict when and where events may occur (e.g., fire breakouts, damaged infrastructure, or people needing rescue).
Moreover, the UAV trajectories vary as they search for events. Therefore, the source UAV and destination (gateway) pairs change with time. 
Combined with high UAV speeds, this results in a highly dynamic network topology, in which existing links break frequently and new links are formed. As a result, the routes between source-destination pairs and the network traffic conditions change often. 
To ensure uninterrupted, reliable, and low-latency communication, a source UAV node should maintain a high-quality multihop route(s) to its gateway node.
Here, `high-quality' routes refer to the longer-lasting and less congested routes, which prevent packet drops due to route breaks and/or congestion build-up.

Existing routing protocols for ground-based mobile ad hoc networks (MANET) are not suitable for decentralized (no supervisory node) UAV networks due to the highly dynamic network topology and frequent link disruptions \cite{mansoor2023fresh, rovira2022review, ateya2019latency, lee2021energy, garg2022accurate, rahmani2022olsr+}. 
Therefore, a number of FANET-specific routing protocols have been discussed in the literature that can be broadly classified as geographic and topology based  \cite{da2021q, mansoor2023fresh, rovira2022review, ateya2019latency, lee2021energy, garg2022accurate,alam2022survey,MCA-OLSR,gharib2022lb,rahmani2022olsr+, BAOMDV}.
Geographic routing protocols, such as \cite{da2021q}, consider a fixed destination node (e.g., base station) whose coordinates are known to all nodes in the network. These schemes are not suitable for the application scenario discussed in this paper, where the network is decentralized and both the source and destination nodes are mobile and can change over time. In contrast, topology-based routing schemes discover and maintain a route to the destination node(s), which makes them better suited for the considered application scenario.

In topology-based routing schemes, nodes recompute the routes periodically (\textbf{proactive} routing schemes) or on-demand (\textbf{reactive} routing schemes) \cite{alam2022survey, rovira2022review}. 
The routing schemes must quickly adapt to topology changes in UAV networks, lest broken routes degrade the flow throughput \cite{alam2022survey}. 
Since high node speeds in FANETs result in frequent changes to network topology, a source node should ideally monitor the route quality in terms of route lifetime ($RLT$), path latency, congestion and route capacity (i.e., data throughput). The selected route should adapt to these varying network and traffic conditions by taking preemptive measures, such as switching to an alternate route (e.g., when the quality of the current route degrades) and decrease congestion along the route.

However, designing a routing protocol with this level of adaptability requires collecting a variety of node and link statistics, such as the number of flows, interference, connectivity to 1-hop neighbors, link capacity, queuing delay, and bit error rates (BER). These time-varying metrics should be updated periodically, as in proactive routing protocols (e.g., optimized link state routing (OLSR) \cite{clausen2003optimized}, and its variants \cite{MCA-OLSR,rahmani2022olsr+,ateya2019latency}). 
However, this increases the control overhead and computational complexity (c.f., \cite{MCA-OLSR}). In addition, proactive routing schemes are vulnerable to security threats, since breaching one node can reveal the entire network topology, including the relative node locations and their IP addresses \cite{saini2020recent}.

On the other hand, reactive routing schemes (e.g., \cite{BAOMDV, li2017lepr,marina2006ad,perkins2003ad,dogra2018q,ahmad2020cache,khudayer2020efficient,perkins2003QoS, jinil2017review, fapojuwo2004perfor, saini2020recent, kumar2013robust, li2019multipath, lee2021energy, choi2019local, hosseinzadeh2024energy, guo2022icra, singhal2022ecms}) 
search routes on-demand, and therefore have low control and computation overhead. A new route is discovered when the current route breaks. 
Since reactive routing schemes do not track changes in the network topology, a source node must either discover a new route (which introduces delay) or risk selecting a sub-optimal route for packet transmission \cite{lau2023aqr}. 

\noindent\textbf{Desired Routing Protocol Characteristics:}
As discussed in \cite{MCA-OLSR}, a routing protocol for decentralized UAV networks should have the following characteristics: (\textit{i}) Find stable route(s) with high throughput to ensure reliable data communication; (\textit{ii}) Introduce low route signaling overhead, delay, and computational complexity; and (\textit{iii}) Monitor the route quality to estimate packet drops due to node mobility and congestion, and take proactive measures to maintain quality. 

We recently designed a proactive routing scheme, called mobility and congestion-aware OLSR (MCA-OLSR) \cite{MCA-OLSR}, for decentralized UAV networks. For the application scenario shown in Fig. 1, source nodes only need to find suitable routes to a limited number of gateway nodes. Here, the use of OLSR protocols would incur higher control (signaling) overhead and route computation complexity in settings with few traffic flows, since each node tries to determine a route to all other nodes in the network.   

This paper presents a novel, hybrid, reactive, ad-hoc on-demand distance vector (Hyd-AODV) routing protocol for decentralized UAV networks to address the issues discussed above. 
This protocol also adopts certain proactive routing features to track limited network topology around the selected route.

\noindent\textbf{Major contributions} of the proposed Hyd-AODV routing protocol include:
\begin{enumerate}
\item\textbf{Multi-metric route selection:} The hop count ($HC$), $RLT$, estimated route latency, and the inter- and intra-flow interference are used to select a stable, longer-lasting and congestion-aware route. 

\item\textbf{Pipe formation:} A 2-hop node region (pipe) is formed around a selected route; it enables a source node to monitor the current route and find a better quality route within the pipe. 

\item Unlike most existing schemes, the proposed routing scheme distinguishes whether a packet may be lost due to a link break or congestion. It addresses congestion buildup through \textit{AoI-based queue management} (discussed next), followed by \textit{preemptive route switching} to select a new, better quality route to both alleviate congestion and avoid broken links, improving the overall quality of service (QoS).
\item\textbf{AoI-based queue management:}
The age-of-information (AoI) \cite{lou2021boosting} refers to the time the packet has spent in the network since it was generated at the source. For latency sensitive applications (e.g., video streaming), a higher AoI value for a packet indicates a lower time-to-expiry. The AoI-based queue management scheme proactively discards packets with high AoI values, if they are not likely to reach the destination node within the allowed latency. The remaining packets are prioritized based on their survivability score (AoI and estimated time-to-destination (ETD)) to maximize their chances of timely delivery.
\item\textbf{Superior performance:} Despite using limited network topology information, the proposed scheme achieves higher throughput, outperforming the AODV, link stability estimation-based preemptive routing (LEPR) \cite{li2017lepr}, 
and OLSR schemes, while incurring few route discoveries and low control overhead and computational complexity. 
\item Compared to existing literature \cite{leonov2018considering, alkhatieb2020performance, kim2023fanet, lau2023aqr, clausen2002comparative}, a more detailed evaluation of the relative performance of reactive and proactive routing schemes is performed for a range of network and traffic settings. 
\end{enumerate}

A customized, discrete event driven simulation framework using ns-3 is developed to evaluate routing performance through extensive simulations, measuring packet delivery ratio (PDR), number of route discoveries, control overhead, and route computation complexity. 

\noindent\textbf{Paper organization:} 
Existing reactive routing schemes are reviewed in Section~\ref{LiteratureReview} for finding a stable, mobility-aware and congestion-free route in FANETs. Next, the proposed Hyd-AODV routing scheme is described in Section~\ref{OurScheme}.
The performance of the proposed scheme is compared with existing reactive and proactive routing schemes in Section~\ref{SimulationSetup&Results}, followed by a discussion of key observations in Section \ref{Keypoints} and conclusions in Section \ref{Conclusion}.

\section{Related Work}
\label{LiteratureReview}
In this section, the existing reactive routing schemes are reviewed, followed by an overview of the AODV \cite{perkins2003ad} and LEPR \cite{li2017lepr} schemes. These two schemes are used as benchmarks in the sequel to evaluate the performance of the proposed Hyd-AODV scheme. 

Traditional reactive routing protocols suffer from the ``broadcast storm'' problem arising from a flood of route request (RREQ) packets in the network \cite{lee2021energy,dogra2018q,ahmad2020cache,choi2019local}. This introduces control overhead and delay due to frequent route rediscoveries in a dynamic network topology. Several methods attempt to ameliorate the problem.  Nodes in \cite{dogra2018q} reduce their RREQ forwarding probability as their queue occupancy increases, reducing the number of congested nodes contesting for the channel to forward control packets. In \cite{ahmad2020cache}, instead of triggering a new network-wide route discovery, the source node requests that the intermediate nodes of the broken route search for an alternate route in their $k$-hop neighborhood to reduce the control overhead. In \cite{khudayer2020efficient}, a node identifies the RREQ forwarding node(s) for each of its different zones based on its neighbor nodes' directions and their local connectivity. However, these schemes \cite{dogra2018q,ahmad2020cache,khudayer2020efficient} can result in poor route selection.

In \cite{choi2019local}, the source and intermediate nodes of the route periodically identify the alternate shortest $HC$ route(s) passing through their 1-hop neighbor nodes. A node switches to a new route when the link with its downstream node breaks due to mobility. This mechanism reduces the number of route discoveries and their resulting delay and control overhead. However, it does not predict the link break and/or congestion buildup, and therefore may transmit packets over an obsolete or congested route. In contrast, the proposed Hyd-AODV scheme proactively switches to a new route before it breaks, and uses multiple metrics (such as $RLT$, route latency and the inter- and intra-flow interference along the route, in addition to $HC$) to select a stable, longer-lasting and less-congested route.

QoS-aware variants \cite{perkins2003QoS, jinil2017review} 
of reactive routing use a subset of statistics which may include congestion, traffic load, delay, node mobility, link stability, signal strength and remaining battery life. For example, nodes in \cite{perkins2003QoS} rebroadcast the RREQ packets only when they can satisfy the flow QoS requirements, such as time-to-live (TTL), minimum bandwidth and jitter. However, this scheme does not track the node mobility and link stability that cause frequent route discoveries in FANETs. In \cite{jinil2017review}, an intermediate node adjusts to the changes in network topology and traffic conditions by reselecting its downstream node 
based on its neighbor distance, latency, load and reliability. However, changing routes locally at intermediate nodes can result in longer, sub-optimal routes \cite{perkins2003ad}. 

The multipath reactive routing schemes \cite{marina2006ad, fapojuwo2004perfor,BAOMDV} find multiple link- or node-disjoint paths at the time of route discovery and use them when the current route quality degrades, decreasing the number of route discoveries \cite{saini2020recent}. However, the quality of the remaining routes may also degrade by the time the primary route breaks, causing flow interruptions and large route switchover overhead and delay \cite{kumar2013robust,li2017lepr}.  Schemes in \cite{kumar2013robust, li2019multipath} prevent the selection of broken routes by periodically monitoring the quality of the remaining routes. A fuzzy-logic based multipath routing scheme is proposed in \cite{lee2021energy} to select a short, low-latency route based on node mobility, residual energy, link quality and stability. It also repairs broken routes using local node information, and initiates a new route discovery if no suitable route is found.

However, the schemes in \cite{kumar2013robust, li2019multipath,lee2021energy} incur large control overhead. To address this issue, a semi-proactive route maintenance mechanism is proposed in LEPR \cite{li2017lepr}, in which the intermediate nodes of the primary route notify the source-destination pair node whenever their link quality degrades. The source node then switches to an alternate route without incurring a large overhead. 

Like the proposed Hyd-AODV scheme, a few other reactive routing schemes (e.g., \cite{hosseinzadeh2024energy, guo2022icra, singhal2022ecms}) also use various route quality metrics for route selection and/or switching. The scheme in \cite{hosseinzadeh2024energy} creates a fixed cylindrical virtual tunnel for a given source-destination pair to limit the number of nodes participating in the route discovery and selection. However, its rigidity prevents it from supporting network topologies that require flexible tunnels.
In \cite{guo2022icra}, a reinforcement-learning (RL)-based strategy is implemented in GCS to dynamically adjust clusters and their heads in response to frequent changes in the network topology. Note that route selection at GCS is not suitable for an autonomous, decentralized UAV network.
An application-aware routing scheme is presented in \cite{singhal2022ecms}, where the source node distributes its traffic load across multiple node-disjoint routes. 
However, unlike Hyd-AODV scheme, it does not preemptively switch to new better-quality routes that may become available due to changes in the network topology or traffic conditions.

LEPR \cite{li2017lepr} is selected as a representative of reactive routing schemes for experimental comparison since:
(\textit{i}) it incorporates link quality in both route selection and switching, and (\textit{ii}) it has been evaluated across a range of network and traffic parameters in the literature, including different node speeds and traffic loads. An overview of the LEPR scheme is given below in Section \ref{LEPRDetails}.

\subsection{Overview of Standard AODV Protocol}
\label{BasicAODV}
AODV \cite{perkins2003ad} is a widely used reactive routing protocol. It uses four types of control packets: RREQ (route request), RREP (route reply), RERR (route error), and Hello packets. The RREQ messages are broadcast in the entire network, while RREP and RERR messages are transmitted using unicast communication, and Hello packets are broadcast in the respective 1-hop neighborhood of each node. 

The source node searches for a route on-demand by flooding the network with RREQ packets. 
An RREQ packet includes the IP addresses of its source and destination nodes, the sequence number of the source node, the last known sequence number of the destination node, and the \textit{Hop Count}. The intermediate node(s) rebroadcasts the RREQ packet if it does not know a route to the destination node. When the destination node receives the first RREQ packet, it transmits an RREP packet towards the source node. The source node then selects the shortest hop count route to the destination node from among the received RREP packets. 

Each node broadcasts its Hello packet after the Hello interval (default value: 1 s). The Hello packet structure is similar to the RREP packet, except the IP address of the RREQ originator node. When a node does not receive any control packet from its 1-hop neighbor node for the Allowed\_Hello\_Loss~$\times$~Hello Interval duration (default value: 3 s), it assumes a link break with that 1-hop neighbor node. When an intermediate node of the route detects a link break, it generates an RERR packet for the source node. Upon receiving an RERR packet, the source node starts a new route discovery.

\subsection{Overview of LEPR Protocol}
\label{LEPRDetails}
The LEPR protocol \cite{li2017lepr} modifies the AODV protocol \cite{perkins2003ad} by computing multiple stable link-disjoint routes during route discovery and proactively switching to an alternative route before the primary route breaks. Its two main components are reactive route discovery and semi-proactive route maintenance. 

To compute the link-disjoint routes, the RREQ and RREP packets carry the information of the first hop and last hop node on the route, respectively. 
Each node computes a link stability metric with each of its 1-hop neighbor nodes by using the locally available information (their link quality, relative distance and mobility) and updates the \textit{link stability metric field} in the RREQ and RREP packets. The path stability of a route is the minimum of all the link stability metrics along the route. Both the source and destination nodes thus know the link-disjoint routes and their stability. 

The source node then selects a route with the highest path stability value and caches the remaining routes. 
When the link stability value at an intermediate node drops below a threshold $k$, the node notifies the destination node by sending a route switch (RSWT) packet. The destination node then transmits a new RREP packet towards the source node on each cached route, which carries the updated path stability metric value of the route. The source node switches to a cached route if its path stability value is greater than a threshold $k'$. Otherwise, it triggers a new route discovery. The value of thresholds $k$ and $k'$ can vary between 0 and 1. 

\textbf{Limitations:}
LEPR \cite{li2017lepr} does not consider the effect of congestion in route selection. Note that nodes in reactive routing schemes rely on the timely reception of their neighbor's control packets; otherwise, a link break is assumed. 
As the number of packet collisions increases with traffic load, a `false' link break can be assumed at a node when it misses its 1-hop neighbor's control packets. In response, the source node either switches to one of its cached routes or initiates a new route discovery. In contrast, the proposed Hyd-AODV scheme incorporates mechanisms to address congestion buildup, including congestion-aware route selection (Section \ref{RREQFlooding}), queue management to postpone congestion buildup (Section \ref{QueueManagement}), and preemptively switching to another high-quality route within the pipe (Section \ref{RouteSwitch}). In addition, each node receives the neighborhood information from other nodes in the pipe, which reduces the probability of flagging a false link break. As a result, the number of route discoveries in Hyd-AODV scheme is much lower compared to LEPR  (see, e.g., Fig. \ref{ch5:ReactiveSchemeCompareRemaining}). 

When the route quality degrades, the source node in LEPR switches to one of its cached routes, which were found during the last route discovery. Although new and better-quality routes may have become available due to network topology changes, LEPR does not find or use them in its semi-proactive route maintenance mechanism. In contrast,  Hyd-AODV scheme proactively finds new better quality routes within the pipe as discussed in Section \ref{RouteSwitch}.

\begin{figure*}[!th]
	\centering
	\includegraphics[width=0.7\linewidth]{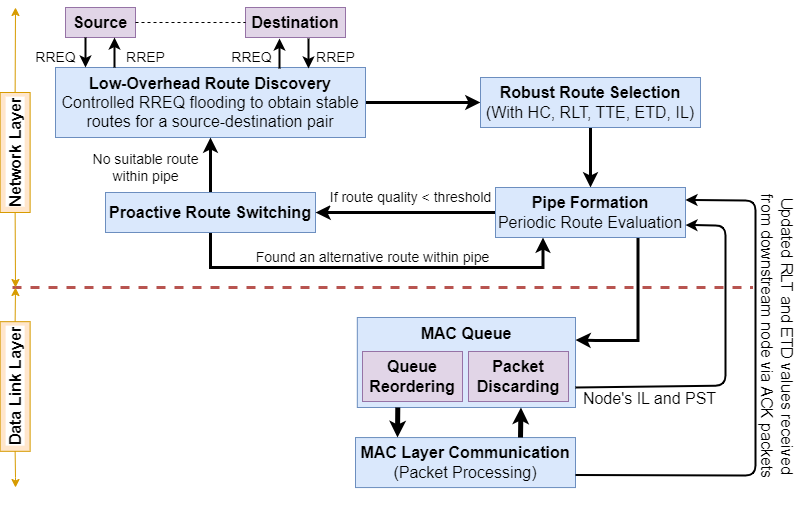}
	\caption {Modules used in the proposed Hyd-AODV routing scheme.}
	\label{ch5:Block-Diagram-Hyd-AODV}
\end{figure*}

\begin{algorithm}[!b]
    \small
	\caption{Overview of Hyd-AODV Scheme} \label{alg:Hyd-AODVRouteSelection}
	\textbf{1. Route discovery:} Source node starts a network-wide route discovery with controlled RREQ flooding (see Section \ref{RREQFlooding})\\
	Destination node sends RREP packet(s), which carries the $HC$, $ETD$ and $IL$ values of the route\\
	\textbf{2. Route selection:} Select a stable and less-congested route $R^*$ from all the received RREPs using (\ref{RouteSelectionConstraints:a}), (\ref{RouteSelectionConstraints:b}) and (\ref{RouteCost})\\
	\textbf{3. Pipe formation:} Intermediate nodes of the selected route $R^*$ collect their 2-hop neighborhood information to create a virtual pipe (see Fig. \ref{ch5:Pipe-Formation-Hyd-AODV})\\
	\textbf{4. Network topology update:} Periodically send the pipe information to the source node\\
	\textbf{5. Route reevaluation and switching:} Source node checks if any of the two conditions of Eq.~(\ref{RouteSwitchConstraints:main}) occurred\\
	\If{\textup{any of the two conditions occurred}}{
	            \eIf{\textup{an alternative route $R$ with quality $>$ threshold exists within the pipe}}{
	                Switch to route R (i.e., $R^*$ = $R$)\\
	                Go to the pipe formation step
	            }{
	                Go to the route discovery step
	            }
			}
\end{algorithm}

\section{Description of Proposed Hyd-AODV Protocol}
\label{OurScheme}
Hyd-AODV routing protocol searches the route(s) on-demand, and then proactively switches to an alternative route before the quality of current route (measured in terms of route longevity and delay using constraints (\ref{RouteSelectionConstraints:a}) and (\ref{RouteSelectionConstraints:b}), respectively, defined in Section III.C) degrades below a threshold. 
It reduces the number of route discoveries, and the resulting overhead and flow interruptions, compared to existing on-demand schemes such as LEPR and standard AODV. 

Fig.~\ref{ch5:Block-Diagram-Hyd-AODV} shows the modules of Hyd-AODV routing scheme, which are discussed below, and summarized in Algorithm \ref{alg:Hyd-AODVRouteSelection}. A source node triggers a route discovery by broadcasting an RREQ message if it does not already have a high quality route to its destination node; details of low-overhead route discovery are given in Section \ref{RREQFlooding}. The destination node responds to the received RREQ messages by transmitting RREP messages towards the source node. Each RREP carries the $HC$, $ETD$ and number of interfering links ($IL$) statistics of the route. 
However, the time-varying metrics (e.g., $RLT$ and $ETD$) are not sufficiently reliable for route selection, as discussed in Section \ref{RoutingMetrics}.
In the proposed scheme, the source node evaluates all the received RREPs through a multi-step route selection mechanism, as discussed in Section \ref{RREQFlooding}. This mechanism first filters available routes based on their quality using the time-varying metrics ($RLT$ and $ETD$). It then selects a low-cost route (see Eq. (\ref{RouteCost})) using the less-sensitive metrics ($HC$ and $IL$).

Theoretical analysis \cite{MCA-OLSR} validated the need for continuous route monitoring and switching when the quality degrades below the desired thresholds, preventing packet drops due to expiry. However, most reactive routing schemes neither proactively track changes in network topology nor trigger route discovery before the current route breaks. To address this limitation, the proposed scheme periodically sends the neighborhood information of the intermediate nodes along the selected route to the source node. Here, the term 'pipe' refers to this neighborhood information. The formation of the pipe, along with analytical and theoretical discussions on the impact of pipe width, are given in Sections \ref{MotivationalAnalysis} and \ref{TheoreticalAnalysis}. 
The source node uses this neighborhood information to reevaluate the current route and switches to an alternative high quality route within the pipe when needed (see proactive route switching details in Section \ref{RouteSwitch}). 
If no route within the pipe meets the required quality, the source node triggers a new route discovery. 

In addition to route switching, nodes should proactively drop packets that are less-likely to reach their destination within the allowed latency. This would free up resources for transmitting the remaining packets that have a higher chance of successful delivery. Therefore, Hyd-AODV scheme incorporates a proactive queue management mechanism, which is discussed in Section \ref{QueueManagement}.

The network modeling and assumptions are discussed in Section \ref{NetworkModeling}, and the modules of Fig. \ref{ch5:Block-Diagram-Hyd-AODV} in Sections \ref{RREQFlooding}--\ref{QueueManagement}, followed by a discussion of control overhead and computational complexity in Section \ref{Control&ComputationalOverhead}.

\subsection{Network Modeling and Assumptions}
\label{NetworkModeling}
A network of low SWaP fixed-wing (FW) UAVs is considered. Since FW-UAVs cannot make sharp turns due to their aerodynamics and high speeds, a smooth-turn (ST) mobility model is used, in which each node independently selects a center and radius, and rotates around the center in a clockwise or counterclockwise direction for a randomly selected duration, until its trajectory changes again \cite{wan2013smooth, garg2022accurate}. 
Here, each UAV is assumed to know its GPS coordinates and is equipped with a low-cost gyroscope, compass, and altimeter. These sensors help determine the orientation, angular velocity, heading direction, and altitude relative to sea level, allowing the UAV to compute its pitch, roll, and yaw values. By tracking these values, the UAV can estimate its recent trajectory, whether it is following a curve or straight line. 
To estimate the center and radius of its trajectory, the UAV uses the last three sensor measurements while flying at a constant altitude. In the case of a straight line trajectory, the radius is assumed to be very large, effectively making the trajectory appear as a straight path. 

Each node broadcasts its trajectory information (i.e., GPS location, center, radius, and movement: clockwise, counter-clockwise, or straight) to its 1-hop neighbors via its Hello messages. 
Based on the possible movement states for a UAV, the following three cases are possible for a UAV pair.
\begin{enumerate}
\item Both UAVs fly in a curve. 
\item One UAV flies in a curve and the other flies in a straight direction.
\item Both UAVs fly in a straight direction at random angles.
\end{enumerate}

\begin{figure}[!th]
	\centering
	\includegraphics[width=\linewidth]{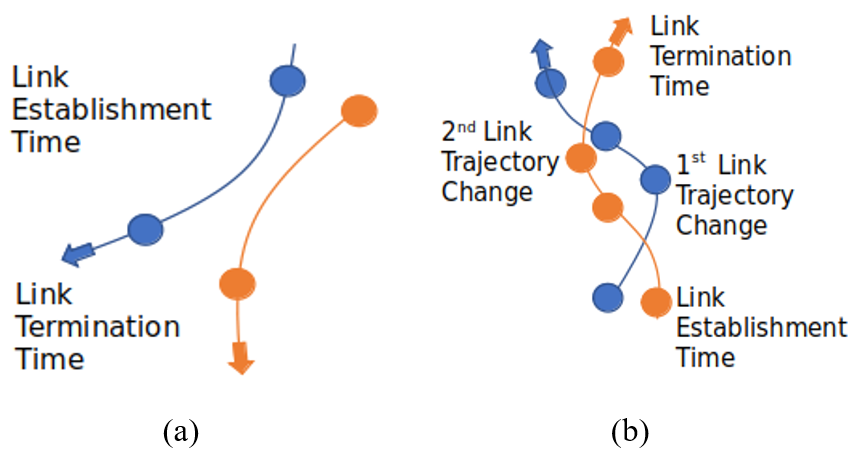}
	\caption {A UAV can randomly change its trajectory, which would change the LLT value of the pair. Therefore, the LLT value is computed when the link is formed or either UAV in the pair changes its trajectory \cite{garg2022accurate,garg2021adaptive}.}
	\label{LLT Recompute}
\end{figure}

A node computes the link lifetime ($LLT$) value for each of its 1-hop neighbor nodes using the mathematical formulation described in \cite{garg2022accurate} for the above three cases.
A UAV pair uses the current trajectory information to compute its $LLT$ value when the link is first established, and updates it when either UAV in the pair changes its trajectory (see Fig. \ref{LLT Recompute} for example). 
In the proposed scheme, the $LLT$ value of a link is included in the Hello message, which is broadcast periodically in the network \cite{MCA-OLSR}.

\subsection{Selection of Routing Metrics}
\label{RoutingMetrics}
Besides $HC$, the received signal strength indicator (RSSI), $RLT$, buffer occupancy ($BO$) and $ETD$ metrics have been used for route cost computation in the literature \cite{ateya2019latency,gharib2022lb,lee2021energy}. However, these routing metrics are not sufficiently reliable. For example, significant interference from neighbor nodes in a dense network can cause inaccurate computation of RSSI values \cite{khudayer2020efficient}. The $RLT$ value can change as the UAVs alter their trajectories \cite{garg2022accurate}, and the values of $BO$ and $ETD$ can increase significantly after the source node starts data transmission along the selected route \cite{MCA-OLSR}.

For latency-sensitive flows with a fixed $TTL$, a high-quality route is quantitatively defined as one that satisfies the following criteria:
\begin{enumerate}
\item $RLT > (TTL - AoI)$: This ensures the route remains operational during the packet’s journey, preventing packet drops caused by route breaks.
\item $ETD << (TTL - AoI)$: This ensures congestion is avoided, preventing packet drops due to excessive delay.
\end{enumerate}
Therefore, like the recent MCA-OLSR scheme \cite{MCA-OLSR}, the Hyd-AODV scheme computes a longer-lasting and low congestion route by using the $HC$ and $IL$ for its route cost computation, and $RLT$ and $ETD$ for route evaluation and switching after data transmission starts on the selected route. An advantage of using the $IL$ metric is that its value for a route changes only when the local topology changes.
As a result, the route quality in Hyd-AODV changes less frequently than existing multi-metric schemes. $HC$ represents the route length in terms of the number of links between the source and destination nodes. 

In Hyd-AODV scheme, each node checks the request-to-send (RTS) and/or data packet received from its neighbors to identify the transmitter-receiver node pair(s) involved in traffic forwarding. It then creates an entry from the transmitter node to receiver node, referred to as an interfering link. The IL metric is calculated as the sum of unique RTS or data packets received by a node within a Hello interval. A node contends for the channel with its interfering links to avoid packet collisions. Consequently, the $IL$ value at a node effectively measures its likelihood of channel access and the congestion in its neighborhood. 
%
The interfering link value for a route ($IL_R$) is the sum of interfering links at all nodes (from the source to destination node) on the route, and is calculated as \cite{ateya2019latency,MCA-OLSR},
\begin{equation}\label{IntfLink}
    IL_{R} = \sum_{j\in N_{R}}\bigg(IL^{\theta}_j + IL^{\phi}_j\bigg),
\end{equation}
where $N_{R}$ is the set of nodes on route $R$ from the source node to the destination. Here, $IL^{\theta}_j$ represents the recent $IL$ value of node $j$ received via its control packet, and $IL^{\phi}_j$ represents the new intra-flow interfering links that will be created at node $j$ when data transmission starts on route $R$. 

The $RLT$ of a route is the time duration after which the route is likely to break. It is the minimum link lifetime, $LLT$, of the links along the route, computed at the source node as \cite{gharib2022lb,MCA-OLSR},
\begin{equation}\label{RLT}
    RLT_{R}=\min_{l \in L_{R}}{(LLT_{l})},
\end{equation}
where $L_{R}$ represents the set of links on route $R$ that connect the source node to the destination via intermediate nodes, and 
the lifetime of each link $l$, $LLT_{l}$, is computed by using the node location (GPS coordinates) and trajectory \cite{garg2022accurate, garg2021adaptive}.

The $ETD$ is the total estimated delay a packet will experience while traveling from the source to destination node on route $R$ \cite{lee2021energy}. It is the sum of the packet service time ($PST$) of each node on the route, where the $PST$ of a node is the total duration for which its packets stay in its medium access control (MAC) queue during the current Hello interval. The $ETD$ value at the source node is computed as \cite{lee2021energy,MCA-OLSR},
\begin{equation}\label{ETD}
    ETD_{R}=\sum_{j\in N_{R}}\bigg(\frac{1}{P_j}\sum_{p\in P_j}{(PST)_p}\bigg)_j,
\end{equation}
where $N_{R}$ is the set of nodes on route $R$  and $P_j$ is the set of data packets successfully transmitted by node $j$ during the previous Hello interval. 

Each node includes its $PST$ and $IL$ values in the Hello packet. An increase in the data generation rate, route length or $IL$ in the neighborhood of a node can cause congestion buildup. This sharply increases the packet queuing delay and reduces its survivability score, which causes packet drops due to expiry \cite{MCA-OLSR}. Therefore, the route quality is tracked using the $HC$, $RLT$, and $ETD$ metrics, and use a preemptive route switching mechanism when the topology changes or congestion buildup reduces the route quality (see Section \ref{RouteSwitch} for details).

\subsection{Low-Overhead Route Discovery}
\label{RREQFlooding}
When a node receives a control packet (e.g., RREQ, RREP or RERR) in the standard AODV protocol, it stores its originator node, destination node and sequence number to avoid reprocessing the same packet. A destination node knows about different routes to the source node through RREQ packets that traverse different routes to reach it. However, it originates an RREP packet only when an RREQ packet is received via a shorter hop route than the previously known route or when the sequence number of the received RREQ packet is higher. 

Instead of blindly forwarding the RREQ packets, a receiver node $i$ in Hyd-AODV scheme broadcasts an RREQ packet only when the following link stability condition is satisfied for transmitter node $j$:
\begin{equation}\label{RREQForwadingCondition}
\centering
 LLT_{i,j}(i)~>~TTL~+~\delta(i)
\end{equation}
Here, $TTL$ represents the time-to-live value of the flow (determined by the application and included in RREQ packets) and $\delta$ is a control parameter, which can be tuned locally at each node depending on its speed and how quickly its neighborhood changes. 

Thus, the RREQ packets traverse towards destination node on longer-lasting routes, reducing the number of control packet transmissions. In order to obtain the \textit{HC}, \textit{ETD} and \textit{IL} statistics of the complete route, only the destination node originates the RREP packet in Hyd-AODV scheme. Note that the \textit{Hop Count} field in an RREP packet represents the hop length of the route from the current node to the destination node. When a node along the reverse route (towards the source node) receives the RREP packet, it adds its own \textit{PST} and \textit{IL} values in the RREP packet's \textit{ETD} and \textit{IL} fields, respectively.

Each RREP packet that a source node receives has traveled on a longer-lasting route due to the constraint imposed by Eq.~(\ref{RREQForwadingCondition}) in RREQ forwarding. Therefore, the residual route lifetime $RLT_R$ of route $R$ used by the RREP packet would be \cite{MCA-OLSR}:
\begin{equation}\label{RouteSelectionConstraints:a}
    RLT_{R}~>~TTL~+~\delta 
\end{equation}

When a source node receives multiple RREPs, it creates a route set $C$ from which it selects the best route using a two-step process discussed below \cite{MCA-OLSR}:

\textbf{Step 1:} Use RREPs that satisfy the following constraint: %
\begin{equation}\label{RouteSelectionConstraints:b}
    \frac{TTE}{ETD_{R}}~\geq~\epsilon_1
\end{equation}

Here, $TTE$ is the time-to-expiry value of the head-of-line (HOL) packet of the flow at the source node at a given time. It is calculated as $(TTL - AoI)$, where $AoI$ denotes the time elapsed since the packet was generated, and $\epsilon_1$ is a constant.

The empirically selected values of $\delta$ and $\epsilon_1$ are the Hello interval and 1.5, respectively. Hence, the route selection mechanism considers only those routes (i.e., $C^* \subseteq C$) which (\textit{i}) will not drop packets due to a link break (Eq. \eqref{RouteSelectionConstraints:a}) and (\textit{ii}) the $TTE$ of packets is $\geq 1.5\times ETD$ (Eq. \eqref{RouteSelectionConstraints:b}).  

\textbf{Step 2:} Compute the cost of each route $R \in C^*$ as shown in \eqref{RouteCost}, and then select the lowest-cost route $R^{*}$:
\begin{equation}\label{RouteCost}
\begin{split}
     Cost_{R} = w_1\bigg(\frac{HC_{R}}{HC_{min}}\bigg) + w_2\bigg(\frac{IL_{R}}{IL_{min}+\alpha\times IL_{R}}\bigg)
\end{split}
\end{equation}
Here, $(\cdot)_{min} = \min\limits_{\forall R~\in~C^*} (\cdot)_{R}$. The scaling factor $\alpha$ is used to convert the $IL$ cost values to the same range as $HC$. The weights of the normalized $HC$ and $IL$ metrics are $w_1$ and $w_2$, respectively, where $w_1+w_2=1$. The values of $w_1=w_2=0.5$ and $\alpha = 0.3$ are used in this paper.

\subsection{Pipe Formation and Proactive Route Switching}
\label{Pipe&Switching}
Due to frequent changes in the network topology, the quality of the current route can degrade with time, while new and better quality routes may become available. In order to improve flow throughput and reduce interruptions, the source node can proactively switch to the highest quality route when the quality of the current route drops below a threshold. However, since the future node trajectories are not known, predicting a high quality route based on the current network topology information is difficult. 
That is why the nodes in proactive routing schemes broadcast their information periodically in the entire network to update the topology information. 
To find a better quality route without high control overhead, the source node collects limited network topology information (the 2-hop neighborhood) around the currently active route, forming a `pipe' that can be used for route switching. 

\subsubsection{Pipe Formation}
\label{pipe_formation}
\begin{figure}[!t]
	\centering
	\includegraphics[width=\linewidth]{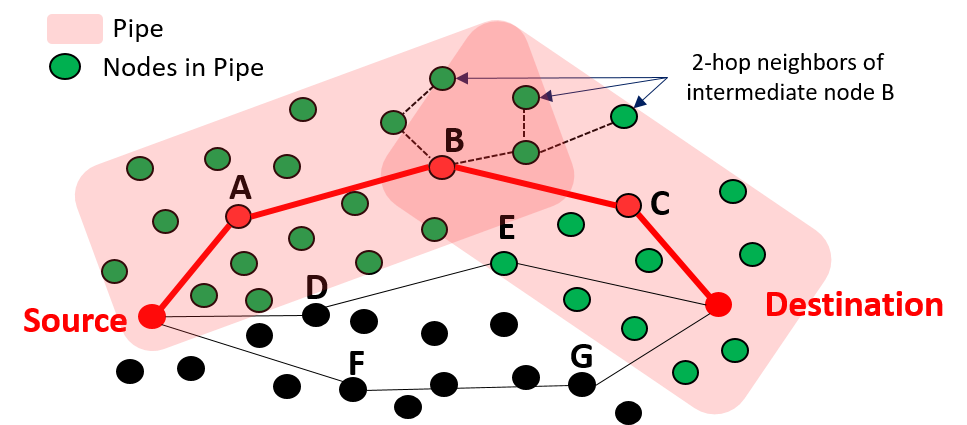}
	\caption {Pipe formation in  Hyd-AODV routing scheme.}
	\label{ch5:Pipe-Formation-Hyd-AODV}
\end{figure}

In typical reactive routing protocols, nodes do not know their 2-hop neighborhoods. Thus, when a node receives a data packet it informs its 1-hop neighbor nodes about its participation in the packet forwarding process by including an `\textit{isActive}' flag in its Hello packet{\footnote{Note that a node can also find its active neighbors by passively listening to the channel.}}. Upon receiving a Hello packet with `\textit{isActive}' set, the node includes its 1-hop neighborhood information (i.e., the nodes' IP addresses and links) in its Hello packet. 
Since the Hello packet includes the sender's GPS coordinates, a node can identify which of its 1-hop neighbors are within transmission range of each other. As a result, the intermediate nodes of the selected route can determine their 2-hop neighborhood information to form the pipe (see Fig. \ref{ch5:Pipe-Formation-Hyd-AODV}). If a node does not receive a data packet within the Active\_Route\_Timeout duration (the default value in the standard AODV protocol is 3 s \cite{perkins2003ad}), it turns off the `\textit{isActive}' flag. Note that the pipe width can also be varied: increasing its width would increase the known fraction of network topology, providing more high-quality route(s) but at the cost of a higher control overhead and computational complexity, as discussed in the sequel (Section \ref{MotivationalAnalysis}).\\

\subsubsection{Sending Pipe Information to the Source Node}
\label{PipeConstruction}
After the destination node receives the first data packet, it schedules an interrupt to send a Notify\_Source control packet to the source node after each Hello interval. When the Notify\_Source packet passes through the intermediate nodes of the route, they append their 2-hop neighborhood information along with the \textit{PST}, \textit{LLT} and \textit{IL} values. If the destination node does not receive a data packet for Active\_Route\_Timeout duration, it suspends the Notify\_Source interrupt.

To reduce the control overhead, an intermediate node constructs a graph using the 2-hop neighborhood information received from its downstream node and includes its own information to compute the cliques \cite{johnston1976cliques}. Then, it includes both the node statistics (\textit{PST} and \textit{IL} values) and link statistics (\textit{LLT} value and \textit{isLinkActive} flag) only once, and forwards to its upstream node.\\

\begin{figure*}[!t]
    \centering
    \includegraphics[width=\linewidth]{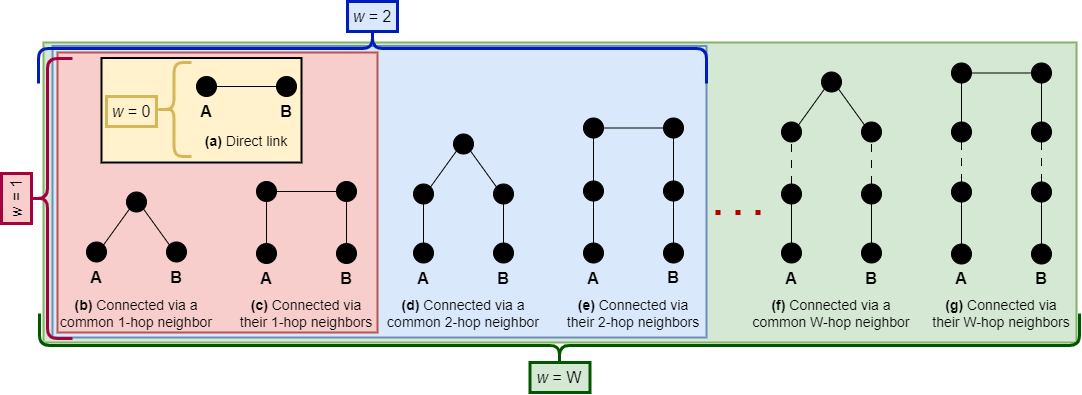}
    \caption{Possible situations to connect two nodes in a network. As the pipe width $W$ increases, more route segments become available to connect the nodes.}
    \label{MA:DifferentScenarios}
\end{figure*}
\begin{table*}[!th]
\caption{Overhead vs. Alternate high-Quality Routes for different Pipe Widths}
\label{TradeOffOverheadVsHQROutes}
\renewcommand{\arraystretch}{1.5}
\resizebox{\linewidth}{!}{%
\begin{tabular}{!{\vrule width 1.5pt}c!{\vrule width 1.5pt}cclcc!{\vrule width 1.5pt}cllclcc!{\vrule width 1.5pt}}
\hline \specialrule{0.1em}{0.1pt}{0.1pt} 
\multirow{3}{*}{\textbf{\begin{tabular}[c]{@{}c@{}}\\ Pipe\\ Width\end{tabular}}} & \multicolumn{5}{c!{\vrule width 1.5pt}}{\textbf{50 Nodes}} & \multicolumn{7}{c!{\vrule width 1.5pt}}{\textbf{100 Nodes}} \\ \cline{2-13} 
 & \multicolumn{3}{c|}{\textbf{Neighbors Tracked*}} & \multicolumn{2}{c!{\vrule width 1.5pt}}{\textbf{No. of High-Quality Routes Found within Pipe}} & \multicolumn{5}{c|}{\textbf{Neighbors Tracked*}} & \multicolumn{2}{c!{\vrule width 1.5pt}}{\textbf{No. of High-Quality Routes Found within Pipe}} \\ \cline{2-13} 
 & \multicolumn{1}{c|}{\textbf{Mean}} & \multicolumn{2}{c|}{\textbf{95\% $CI^{\#}$}} & \multicolumn{1}{c|}{\textbf{\begin{tabular}[c]{@{}c@{}}New Route Length $\leq$ \\ Current Route Length\end{tabular}}} & \textbf{\begin{tabular}[c]{@{}c@{}}New Route Length \textgreater \\ Current Route Length\end{tabular}} & \multicolumn{3}{c|}{\textbf{Mean}} & \multicolumn{2}{c|}{\textbf{95\% $CI^{\#}$}} & \multicolumn{1}{c|}{\textbf{\begin{tabular}[c]{@{}c@{}}New Route Length $\leq$ \\  Current Route Length\end{tabular}}} & \textbf{\begin{tabular}[c]{@{}c@{}}New Route Length \textgreater \\ Current Route Length\end{tabular}} \\ \hline\specialrule{0.1em}{0.1pt}{0.1pt} 
\textbf{1} & \multicolumn{1}{c|}{5} & \multicolumn{2}{c|}{{[}4.7, 5.2{]}} & \multicolumn{1}{c|}{1} & 3 & \multicolumn{3}{c|}{8} & \multicolumn{2}{c|}{{[}7.2, 8.6{]}} & \multicolumn{1}{c|}{5} & 13 \\ \hline
\textbf{2} & \multicolumn{1}{c|}{9} & \multicolumn{2}{c|}{{[}8.6, 9.8{]}} & \multicolumn{1}{c|}{4} & 6 & \multicolumn{3}{c|}{17} & \multicolumn{2}{c|}{{[}15.2, 19.1{]}} & \multicolumn{1}{c|}{13} & 15 \\ \hline
\textbf{3} & \multicolumn{1}{c|}{13} & \multicolumn{2}{c|}{{[}11.5, 13.6{]}} & \multicolumn{1}{c|}{14} & 18 & \multicolumn{3}{c|}{28} & \multicolumn{2}{c|}{{[}26.1, 30.5{]}} & \multicolumn{1}{c|}{45} &79 \\ \hline
\textbf{4} & \multicolumn{1}{c|}{16} & \multicolumn{2}{c|}{{[}14.4, 17.4{]}} & \multicolumn{1}{c|}{24} & 22 & \multicolumn{3}{c|}{43} & \multicolumn{2}{c|}{[39.8, 46.3]} & \multicolumn{1}{c|}{71} & 92 \\ \hline
\specialrule{0.1em}{0.1pt}{0.1pt} 
\end{tabular}
}\newline\newline
\textbf{*} Shows total neighbors tracked per node. \;\;$CI^\textbf{\#}$ represents confidence interval.
\end{table*}

\subsubsection{Analysis of Pipe Width}
\label{MotivationalAnalysis}
Fig. \ref{MA:DifferentScenarios} shows different scenarios which connect two nodes A and B on a route, either directly or through their neighbors, for different pipe widths $W$. Many existing reactive routing schemes \cite{choi2019local,kumar2013robust,saini2020recent,jinil2017review} use only the 1-hop neighborhood information to locally repair a broken link. If the 1-hop neighbors cannot find an alternative route to the downstream node, a new route discovery is initiated. 
As shown in Fig. \ref{MA:DifferentScenarios}, more route segments can become available for local link repair when a node collects $W$-hop neighborhood information. 
For example, Fig. \ref{MA:DifferentScenarios}(a)-(e) shows five possibilities for connecting nodes A and B when $W=2$. 
However, simply repairing the link using this connection may increase the route length significantly: for example, route length can increase by up to 2$W$ to repair link A-B in Fig. \ref{MA:DifferentScenarios}(g). 
To avoid unnecessary increase in route length, the source node forms a new route within the pipe in Hyd-AODV scheme. 

As shown in Table \ref{TradeOffOverheadVsHQROutes}, both the overhead (measured by the number of neighbors tracked by each node) and the number of high-quality routes available within the pipe increase significantly with pipe width. A source node can obtain on average four or more routes within the pipe when $W=1$. 
However, due to limited neighborhood information, these routes often include edge nodes, which can cause link breaks at high node speeds. 
A decrease in the number of new route discoveries (initiated when a high quality route cannot be found within the pipe) is observed for a higher pipe width. Therefore, $W=2$ is selected that provides sufficient high-quality routes while keeping the overhead low.\\

\subsubsection{Theoretical Analysis for Pipe Width}
\label{TheoreticalAnalysis}
The impact of pipe width on traffic load and the mechanisms proposed in this work (including route selection, switching, and queue management) is analyzed. Then theoretical analysis from \cite{MCA-OLSR} is used to examine the trade-offs associated with pipe width for these mechanisms.

In a dynamic network with random source-destination pairs, the total number of nodes involved in traffic forwarding typically increases with the number of flows. This also increases the number of common nodes that either serve multiple routes or fall within the transmission range of intermediate nodes on these routes. When nodes are uniformly distributed, the number of such common nodes tracked within the pipe grows with pipe width, as shown in Table \ref{TradeOffOverheadVsHQROutes}. These common nodes, when subject to mutual interference, create interfering links that influence the $IL$ metric. Thus, the relationship between traffic flows and pipe width can be observed through the $IL$ metric.

On the other hand, factors like link breaks, channel fading, and packet collisions contribute to congestion, which increases queuing delay. High traffic loads—caused by a high data generation rate or multiple traffic flows—further exacerbate this delay. For latency-sensitive flows, increased queuing delay can lead to packet drops due to expiry, which reduces flow throughput and impacts metrics such as $PST$, $ETD$, and the packet survivability score, computed as $\frac{(TTL-AoI)}{ETD}$ (see Section \ref{QueueManagement} for details).
To mitigate congestion and reduce queuing delay, the source node should switch to an alternative route and drop the packets that are less-likely to reach their destination. Since pipe width influences the number of available routes within the pipe (see Table \ref{TradeOffOverheadVsHQROutes}) and their lengths (see Fig. \ref{MA:DifferentScenarios}), it directly affects route selection, switching, and queue management.

Therefore, the discussion below considers how flow rate, $HC$, and $IL$ affect packet service rate, queuing delay and packet survivability score, and the role of pipe width in improving these metrics and enhancing the proposed mechanisms. A well-chosen pipe width ensures a sufficient number of alternative routes while decreasing interference to neighboring flows, number of route discoveries triggered, and overhead incurred in collecting the neighborhood information.

A G/M/1/$\infty$ queuing system is considered, where each source node generates packets at a constant rate (i.e., mean packet arrival rate) $\lambda$, stores them in FIFO order, and transmits one packet at a time when the channel is available \cite{QueuingTheory}. Here, the queuing delays of the packets are independent identically distributed random variables. A constant $TTL$ value is assumed for each packet. A node drops the packet from its queue when the packet $TTE$ value is 0. As a result, the packets will not be dropped due to buffer overflow when the node has a finite but large buffer. Therefore, the simulation setup used in this paper is similar to the queuing system with $\infty$ buffer size. 

The average $PST$ of packets of a node $i$ in such system is $\frac{1}{(\mu_i-\lambda)}$, where the average packet service rate $\mu$ of node $i$ is computed as \cite{QueuingTheory},
\begin{equation}\label{ServieRate}
   \mu_i~=~\frac{Q_i}{S\times IL_i}
\end{equation}
Here, $Q_i$ is the channel rate (in bits per s) at node $i$ and $S$ is the packet size (in bits). The value of $\mu_i$ decreases as the number of interfering links ($IL_i$) in the neighborhood of node $i$ increases. Similar network and traffic conditions are assumed at each node. Hence, $Q_i = Q$, $IL_i = IL$, and $\mu_i = \mu;~\forall i \in V$, where $V$ is the set of nodes in the network. 

If the average route length is $H$ hops, the average time-to-destination (ATD) for a packet is $\frac{H}{(\mu-\lambda)}$. The average $TTE$ of a packet at a node can be computed as \cite{QueuingTheory},
\begin{equation}
    TTE = TTL - U\times n,~
    where~U = \frac{\lambda}{\mu(\mu-\lambda)}
\end{equation}
Here, $U$ represents the average queuing delay of a packet and $n$ is the distance in hops from the source node. 

\begin{figure}[!t]
\centering
\includegraphics[width=0.9\linewidth]{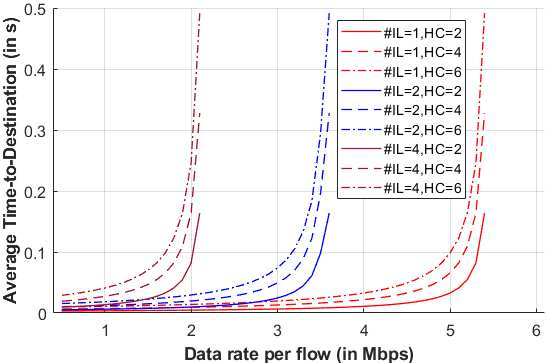}
\caption{Theoretical value of average time-to-destination for different data rate, $IL$ and $HC$ values \cite{MCA-OLSR}.}
\label{ETDvsDataRate}
\end{figure}

\begin{figure}[!t]
\centering
\includegraphics[width=0.9\linewidth]{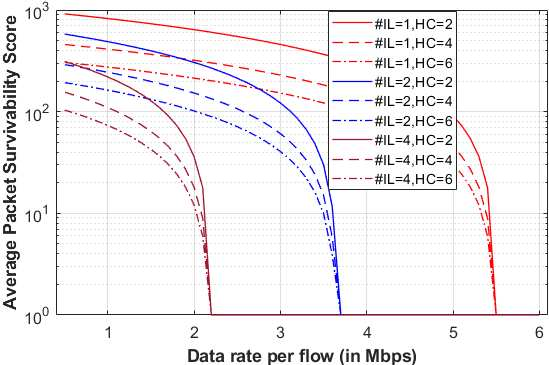}
\caption{Theoretical value of packet survivability score for different data rate, $IL$ and $HC$ values \cite{MCA-OLSR}.}
\label{SurvivalRatevsDataRate}
\end{figure}

An increase in the data generation rate and/or $IL$ in the neighborhood of a node results in $\lambda>\mu$ condition. Note that a node reaches this condition at a lower value of $\mu$ in a fading channel. When this condition occurs, the congestion builds up at the node, which sharply increases the packet queuing delay and $PST$. As a result, the ATD also increases sharply\footnote{The impact on ETD will be the same as that on ATD.} in Fig. \ref{ETDvsDataRate}. This quickly reduces the packet survivability score (see Fig. \ref{SurvivalRatevsDataRate}), which causes packet drops due to expiry. If the source node has alternative routes within the pipe, it can switch to one with a lower $IL$ value to improve both $\mu$ and ATD. Ideally, the new route should lie outside the direct transmission range of nodes on the previous routes to minimize interference, which often results in a longer $HC$ route. Otherwise, the packet survivability score will remain low, necessitating the use of queue management mechanism to proactively drop packets and conserve resources for packets that have a higher likelihood of reaching the destination node.

This requires a pipe width of two or more as shown in Fig. \ref{MA:DifferentScenarios} and validated by simulation in Table \ref{TradeOffOverheadVsHQROutes}. However, an increase in route length leads to a higher ATD (see Fig. \ref{ETDvsDataRate}) and a lower packet survivability score (see Fig. \ref{SurvivalRatevsDataRate}). Therefore, excessively long routes are also not desirable. At the same time, a lower pipe width limits the scope of pipe boundary, which can decrease the interference on neighboring flows. This, as a result, decreases the $IL$ value of the route and provides low-cost routes for selection.
Therefore, the pipe width of two is chosen in this paper, as it offers sufficient alternative routes outside the transmission range of the previous route while keeping the overhead of collecting neighborhood information low (as discussed in Section \ref{MotivationalAnalysis}).


\subsubsection{Route Switching Mechanism}
\label{RouteSwitch}
When the quality of the current route {$R^*$} degrades below a threshold (corresponding to Constraints 
(\ref{RouteSwitchConstraints:a}) and (\ref{RouteSwitchConstraints:b})), the source node uses a breadth first search (BFS) algorithm to find all routes within the pipe \cite{MCA-OLSR}. 
\begin{subequations}\label{RouteSwitchConstraints:main}
\begin{align}
 RLT_{{R}^{*}}~&<~TTE~+~\delta \label{RouteSwitchConstraints:a}\\
 \frac{TTE}{ETD_{R^{*}}}~&<~\epsilon_2\label{RouteSwitchConstraints:b}
\end{align}
\end{subequations}
The conditions \eqref{RouteSwitchConstraints:a} and \eqref{RouteSwitchConstraints:b} correspond to violations of the route selection 
constraints \eqref{RouteSelectionConstraints:a} and \eqref{RouteSelectionConstraints:b}. The selected route $R^{*}$ is used as long as the packet $TTE \geq \epsilon_2\times ETD$. The empirically selected value of $\epsilon_2$ is 1.1.
If a new better-quality route is found, the source node switches to the new route; otherwise it triggers a new route discovery. 
When the source node switches to a new route, the existing pipe is abandoned and a new pipe is formed around the new route. The route switching mechanism, therefore, reduces the number of route discoveries and the resulting discovery overhead and delay. It also minimizes the interruption in flow throughput. \\

\subsection{AoI-Aware Queue Management}
\label{QueueManagement}
The congestion at a node can increase due to a high data rate, inter- and intra-flow interference, and frequent link breaks, which increase the number of packets in the queue and their queuing delay. To mitigate these issues, each node in Hyd-AODV scheme periodically reevaluates its queue using an AoI-based queue management policy \cite{MCA-OLSR}, where AoI measures the time each packet has spent in the network:
\begin{enumerate}
\item[(\textit{i})] Instead of the FIFO order, rearrange the packets of the queue at a node in ascending order of their survivability score, $\frac{(TTL-AoI)}{ETD}$.
\item[(\textit{ii})] Drop any packets with a survivability score lower than a threshold (e.g., 0.7), as they are not likely to be delivered to the destination before their expiry.
\end{enumerate}
This prioritizes sending packets with lower (but not too low) survivability scores, increasing their chances of reaching the destination node before their TTL expiry.

\subsection{Control Overhead and Computational Complexity}
\label{Control&ComputationalOverhead}
\subsubsection{Control Overhead}
In addition to the information sent in the RREQ, RREP, RERR and Hello packets in the standard AODV protocol, each node in the proposed Hyd-AODV scheme includes the following information in its Hello packet:
\begin{itemize}
    \item Its GPS location (using 6 bytes for ($x$,$y$,$z$) coordinates) and trajectory information, which includes the center coordinates (6 bytes), radius and node movement direction in the ST mobility model (2 bytes) \cite{wan2013smooth}. Note that a node sends its trajectory information only when it forms a new link or changes its current trajectory.
    \item Its $PST$ and $IL$ values (up to 2 bytes). 
    \item If the node is a 1-hop neighbor of an active node on the route,  it broadcasts the $LLT$, $PST$, $IL$ and $isLinkActive$ values for each of its 1-hop neighbor nodes (variable size). 
\end{itemize}

In addition, the destination node periodically sends a Notify\_Source packet to the source node, which carries the 2-hop neighborhood information (i.e., node IP address and their $PST$, $IL$ and $LLT$ values) of all intermediate nodes on the route (variable size).\\

\subsubsection{Route Computation Complexity}
\label{CompComplexity}
In this scheme, a source node finds all routes to the destination node within the pipe using the BFS algorithm, which has a worst time complexity of $O(T {V_p}^2 (E_p+V_p))$, where $T$ is the number of times routes are recomputed, and $V_p$ and $E_p$ are the number of nodes and links, respectively, within the pipe. In addition, the intermediate nodes compute the cliques of their 2-hop neighborhood using the Bron-Kerbosch algorithm \cite{johnston1976cliques}, which has a worst time complexity of $O(3^{V_p/3})$. Note that the clique computation is optional but helps in reducing the control overhead. 

\begin{figure*}[!b]
	\centering
	\includegraphics[width=0.8\linewidth]{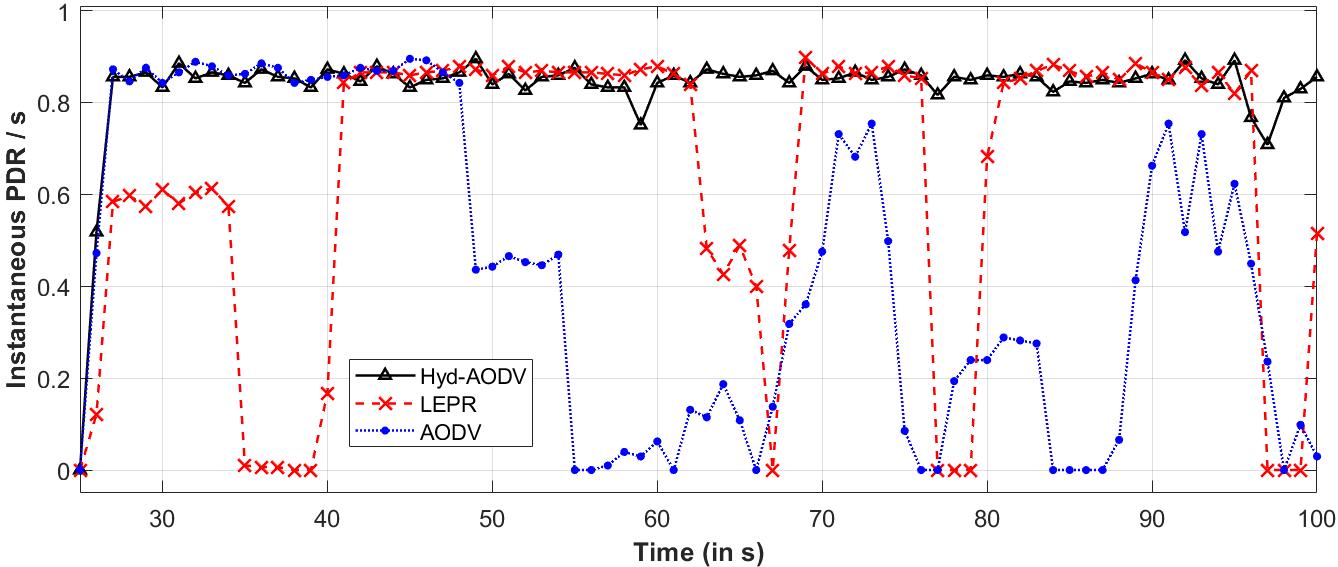}
	\caption{Instantaneous PDR for Hyd-AODV, LEPR and AODV schemes.}
	\label{FigQoSCompare}
\end{figure*}

\section{Simulation Setup and Results}
\label{SimulationSetup&Results}
A customized discrete event driven simulation framework, using ns-3 version 3.29, is developed to implement Hyd-AODV scheme{\footnote{The code is available \href{https://drive.google.com/drive/folders/1vTJNTnlodyVdzZfaONmhMVWlkW1ulohf}{\underline{here}}.}} The simulation and network parameters are summarized in Table \ref{AODVSimulationParameters}. In the simulation, fixed-wing, low SWaP UAVs fly in an 8$\times$8 km$^2$ area (in which fixed communication infrastructure, such as a cellular network, is not available) following the ST mobility model of  \cite{wan2013smooth, garg2022accurate, MCA-OLSR, kim2023fanet, garg2021adaptive}. The UAVs fly in the X-Y plane at a constant altitude, about 1 km above the ground \cite{switchblade, coyote} and perform collision avoidance through trajectory modifications \cite{collision_avoidance}. 
Performance is evaluated at multiple network settings, including node density (50 and 100 UAVs), low to high node speeds (20 m/s and 50 m/s), and low to high traffic loads (40 kbps to 3 Mbps bit rate). 

The UAVs are equipped with GPS and have a signal transmission range of 1 km (e.g., \cite{MCA-OLSR, oubbati2021dispatch, ieee80211ah}). Since there are no reflections or obstacles in the air, a free space line-of-sight propagation with path loss is used for the air-to-air (A2A) channel between UAVs (e.g., \cite{ieee80211ah, oubbati2021dispatch}). The packet size and TTL values are 1 kB and 3 s, respectively, and the MAC queue stores up to 1000 packets. The carrier sense multiple access with collision avoidance (CSMA/CA) protocol is used in the MAC layer.
Each simulation is run for 600 s and each experiment is repeated 40 times, with the source-destination pair(s) of a flow selected randomly in each run. 

\begin{table}[!t]
\centering
\caption{Simulation Parameters}
\label{AODVSimulationParameters}
\renewcommand{\arraystretch}{1.5}
\resizebox{0.7\linewidth}{!}{%
\begin{tabular}{|c|c|}
\hline
\textbf{Parameters}     & \textbf{Values}        \\ \hline
Simulation Area         & $8\times~8~km^2$                \\ \hline
Channel Rate            & 11 Mbps                \\ \hline
Transmission Range      & 1 km                   \\ \hline
Node Density            & {50, 100}     \\ \hline
Node Speed (in m/s)     & {20, 50}   \\ \hline
Number of Traffic Flows & {1, 3, 10} \\ \hline
Flow Rate               & 40 kbps to 3 Mbps    \\ \hline
Packet Size             & 1 kB                   \\ \hline
Time-to-Live (TTL)      & 3 s                    \\ \hline
\end{tabular}%
}
\end{table}

The performance of the proposed Hyd-AODV scheme is compared with the standard AODV and LEPR schemes in Section \ref{CompareWithReactiveScheme}, followed by proactive routing schemes (i.e., standard OLSR and MCA-OLSR) in Section \ref{CompareWithProactiveScheme}. 

\subsection{Performance Metrics}
\label{PerformanceMetrics}
The following performance metrics are used:
\begin{enumerate}
\item\textbf{Packet delivery ratio (PDR)} of a flow is the ratio of total data packets received at the destination node to the total data packets generated at the source node. 
The \emph{flow throughput} can be computed as PDR $\times$ data rate, while the \emph{packet loss ratio} equals (1-PDR). When plotting the average PDR, the 95\% confidence interval is also indicated (shaded region).

\item\textbf{Number of routes computed} is the total number of routes computed during the simulation.

\item\textbf{Number of route discoveries} is the number of RREQ packets generated per flow at the source node. A lower value signifies a more stable route(s). 

\item\textbf{Number of route control packets} is the total number of route setup, update and maintenance packets (i.e., RREQ, RREP, RERR and Hello) transmitted over the network during the simulation duration. This metric also includes the Notify\_Source packets in Hyd-AODV scheme and RSWT packets in LEPR scheme. 

\item\textbf{Control (signaling) overhead} is the total size (in MB) of the route control packets transmitted during the simulation duration. 
\end{enumerate}

Experimental results are averaged over 40 runs of 600 s duration each.

\subsection{Comparison with Reactive Routing Schemes}
\label{CompareWithReactiveScheme}
Here, the performance of  Hyd-AODV scheme is compared with two reactive routing schemes (AODV and LEPR) in terms of PDR, number of route discovery, and route control overhead. \\
\begin{figure*}[!b]
	\centering
	\includegraphics[width=0.9\linewidth]{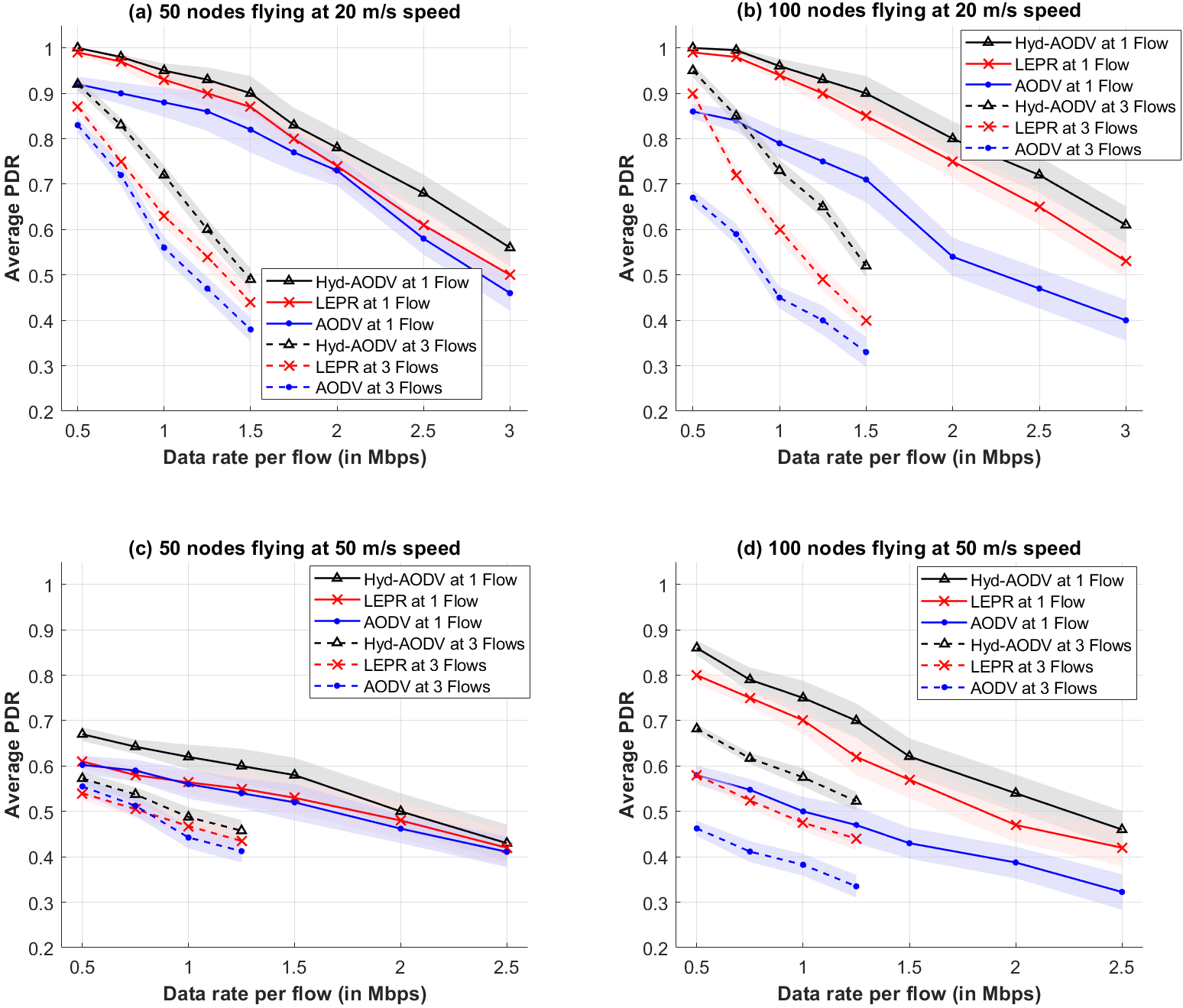}
	\caption{Average PDR for Hyd-AODV, LEPR and AODV schemes for different number of data flows and data rates, at varying node densities and speeds.}
	\label{ch5:ReactiveSchemeCompareAll}
\end{figure*}

\begin{table*}[!b]
\centering
\caption{Average PDR for 10 Low Data Rate Flows}
\label{10Flows}
\renewcommand{\arraystretch}{1.5}
\resizebox{0.9\linewidth}{!}{%
\begin{tabular}{!{\vrule width 1.5pt}c!{\vrule width 1.5pt}c|c|c|c|c!{\vrule width 1.5pt}}
\hline
\specialrule{0.1em}{0.1pt}{0.1pt} 
\textbf{Data Rate per Flow} &
  \textbf{Scheme} &
  \textbf{50 Nodes @ 20 m/s} &
  \textbf{100 Nodes @ 20 m/s} &
  \textbf{50 Nodes @ 50 m/s} &
  \textbf{100 Nodes @ 50 m/s} \\ \hline
\specialrule{0.1em}{0.1pt}{0.1pt} 
\multirow{3}{*}{\textbf{40 kbps}}  & \textbf{AODV} & 0.92 & 0.88 & 0.56 & 0.55 \\ \cline{2-6} 
                             & \textbf{LEPR} & 0.92 & 0.88 & 0.56 & 0.65 \\ \cline{2-6} 
                             & \textbf{Hyd-AODV} & 0.95 & 0.99 & 0.6  & 0.75 \\
                             \cline{2-6} 
                             & \textbf{OLSR} & 0.85 & 0.85 & 0.37  & 0.42 \\
                             \cline{2-6} 
                             & \textbf{MCA-OLSR} & 0.97 & 1 & 0.56  & 0.81 \\
                             \hline
\specialrule{0.1em}{0.1pt}{0.1pt} 
\multirow{3}{*}{\textbf{200 kbps}} & \textbf{AODV} & 0.85 & 0.7 & 0.47 & 0.44 \\ \cline{2-6} 
                             & \textbf{LEPR}     & 0.87 & 0.79 & 0.47 & 0.48 \\ \cline{2-6} 
                             & \textbf{Hyd-AODV} & 0.92 & 0.93 & 0.52 & 0.64 \\ 
                             \cline{2-6} 
                             & \textbf{OLSR} & 0.78 & 0.81 & 0.33  & 0.39 \\
                             \cline{2-6} 
                             & \textbf{MCA-OLSR} & 0.95 & 0.99 & 0.55  & 0.77 \\
                             \hline                          
\specialrule{0.1em}{0.1pt}{0.1pt} 
\end{tabular}}
\end{table*}

\subsubsection{PDR Performance Comparison}
First, the instantaneous PDR is studied. Then, the impact of number of data flows and data rate on the average PDR performance is discussed, followed by the impact of node densities and speeds.

\paragraph{Instantaneous PDR Performance}
Fig. \ref{FigQoSCompare} shows the instantaneous PDR values for each scheme for 100 nodes flying at 20 m/s speed and a data flow rate of 2.5 Mbps. Performance is shown for the first 100 s, where the initial 25 s are used for network stabilization. Unlike AODV and LEPR, Hyd-AODV scheme provides a consistently higher instantaneous PDR because it quickly adapts to dynamic network topology by predicting the congestion buildup and link breaks. As a result, it can provide uninterrupted communication throughout the simulation duration despite a dynamic network topology. 
Although not shown, the instantaneous PDR plots for other simulation settings also showed superior performance over AODV and LEPR. 

\paragraph{Impact of Traffic Rates}
Fig. \ref{ch5:ReactiveSchemeCompareAll} shows the average PDR of each scheme for one and three flows across varying data rates until the network becomes congested. The PDRs of all the schemes decrease as the traffic load increases.
Since Hyd-AODV scheme incurs a lower RREQ flooding overhead, selects stable and congestion-free (or less congested) routes, and preemptively switches to a new route when the quality of current route degrades, it provides a higher average PDR than the standard AODV and LEPR schemes. 

For 50 nodes at 20 m/s speed, Hyd-AODV achieves up to 29\% and 14\% higher PDR compared to the AODV and LEPR schemes, respectively, in Fig. \ref{ch5:ReactiveSchemeCompareAll}(a). 
For a higher node density (100 nodes) at 20 m/s, Hyd-AODV achieves up to 63\% and 33\% higher PDR compared to AODV and LEPR schemes, respectively (see Fig. \ref{ch5:ReactiveSchemeCompareAll}(b)). Although more routes become available for a source-destination pair in the higher density network, AODV continues to use the current route until a new route discovery is triggered. 
Since a source node tracks the nodes within the pipe around the active route (in Hyd-AODV) or caches different link-disjoint routes (in LEPR), it switches to a less-congested route, whenever needed. Therefore, the PDR performance of Hyd-AODV and LEPR is improved.  

Fig. \ref{ch5:ReactiveSchemeCompareAll}(c) shows the PDR for 50 nodes flying at 50 m/s speed. Since the links break more frequently at 50 m/s, which require a higher number of route discoveries, the PDR performance degrades for all three schemes as compared to node speed of 20 m/s in Fig. \ref{ch5:ReactiveSchemeCompareAll}(a). However, PDR performance of Hyd-AODV is still up to 12\% and 11\% higher compared to AODV and LEPR schemes, respectively.

Fig. \ref{ch5:ReactiveSchemeCompareAll}(d) shows the PDR performance for 100 nodes at 50 m/s. The PDR performance of Hyd-AODV is up to 56\% and 21\% higher than AODV and LEPR, respectively. Although the route discoveries increase at 50 m/s, the number of available routes also increases with an increase in the node density. As a result, the source node in both Hyd-AODV and LEPR schemes successfully switches to an alternate route before the primary route breaks. Therefore, the PDR values in both these schemes improve at 100 nodes as compared to 50 nodes. 
However, as discussed earlier in this subsection, Hyd-AODV scheme incurs a lower RREQ overhead, selects stable and congestion-free (or less congested) routes, and preemptively switches to a new route when the quality of current route degrades and provides a higher average PDR than LEPR scheme.

\begin{figure*}[!t]
	\centering
	\includegraphics[width=0.7\linewidth]{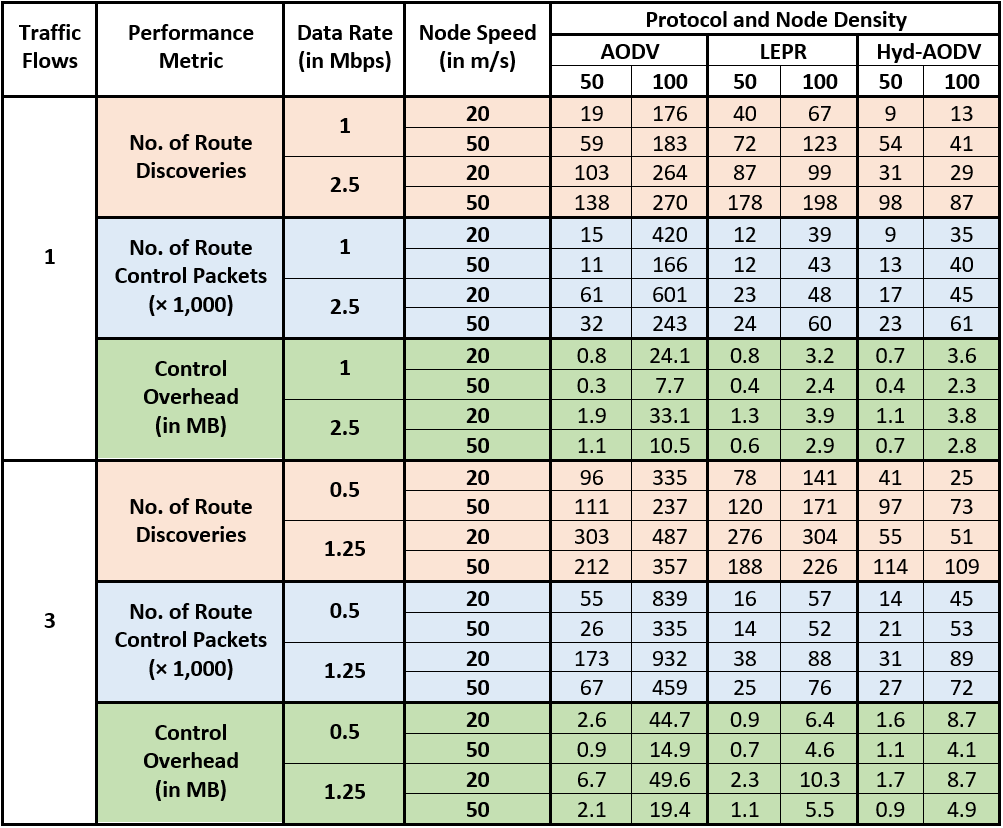}
	\caption{Performance comparison of Hyd-AODV, LEPR and AODV schemes in terms of average route discoveries, route control packets and control overhead at different traffic loads, node speeds, and densities.}
	\label{ch5:ReactiveSchemeCompareRemaining}
\end{figure*}

\paragraph{Impact of Node Density}
When the node density increases from 50 to 100 nodes, the PDR performance of Hyd-AODV and LEPR schemes at 50 m/s improves by up to 28\% and 31\%, respectively. Although routes break more frequently at 50 m/s than at 20 m/s, the fact that more routes are available at higher node density enables both schemes to switch to an alternate route when the current route breaks.

\paragraph{Impact of Node Speed}
At higher speed, the network topology changes faster and creates more frequent route breaks, decreasing the PDR performance. The Hyd-AODV scheme selects a longer-lasting and congestion-free (or less-congested) route, and performs route switching within the pipe when the current route degrades. It thus spends less time finding a new route as compared to the AODV and LEPR schemes. 
AODV does not monitor the route quality and so needs extra time to detect a route break;
on the other hand, LEPR evaluates the stability of all cached routes before initiating a new route discovery, which incurs delay when links break frequently. Thus, Hyd-AODV achieves up to 58\% and 21\% higher PDR compared to AODV and LEPR, respectively, for three flows at the higher speed of 50 m/s in Fig. \ref{ch5:ReactiveSchemeCompareAll}(d).

\paragraph{Performance for 10 Sensor Data Flows} Table \ref{10Flows} shows the PDR performance comparison for 10 sensor data flows at two different data rates for two node densities and two speeds. Here, the per flow data rates are 40 kbps and 200 kbps, which correspond to a total of 400 kbps and 2 Mbps data traffic generated per second, respectively. The Hyd-AODV scheme achieves higher PDR as compared to the AODV and LEPR schemes at each setting. 
Since the number of alternative routes formed within the pipe for each traffic flow increases at a higher node density, the PDR performance of Hyd-AODV scheme improves significantly when the node density increases from 50 to 100 nodes at node speed of 50 m/s.\\

\subsubsection{Comparing the Number of Route Discoveries}
\label{RouteDiscoveries}
Fig. \ref{ch5:ReactiveSchemeCompareRemaining} shows the number of route discoveries in the three reactive routing schemes for one and three traffic flows at two data rates, and two different node speeds and densities. The Hyd-AODV scheme has fewer route discoveries than AODV and LEPR, because it searches for an alternate route within the pipe before initiating a new route discovery. 

The route discoveries increase with traffic load.
Since the network becomes congested and the packet collisions and channel access time increase at higher traffic loads, the Hello and other control packets are delayed. When a node does not hear from its 1-hop neighbor node(s) within a set duration, it assumes a link break, which can trigger a new route discovery in AODV or route switching in LEPR. 
In Hyd-AODV, each intermediate node on the route receives the neighborhood information from all of its neighbor nodes within the pipe. This increases redundancy in the received information, which reduces the probability of falsely flagging a link break and prevents triggering route discovery or switching in Hyd-AODV when the network is congested.

The number of route discoveries increases in both AODV and LEPR as the node density increases from 50 to 100 nodes, but decreases for Hyd-AODV scheme, since more alternate routes become available within the pipe at a higher node density.\\

\begin{figure*}[!b]
	\centering
	\includegraphics[width=\linewidth]{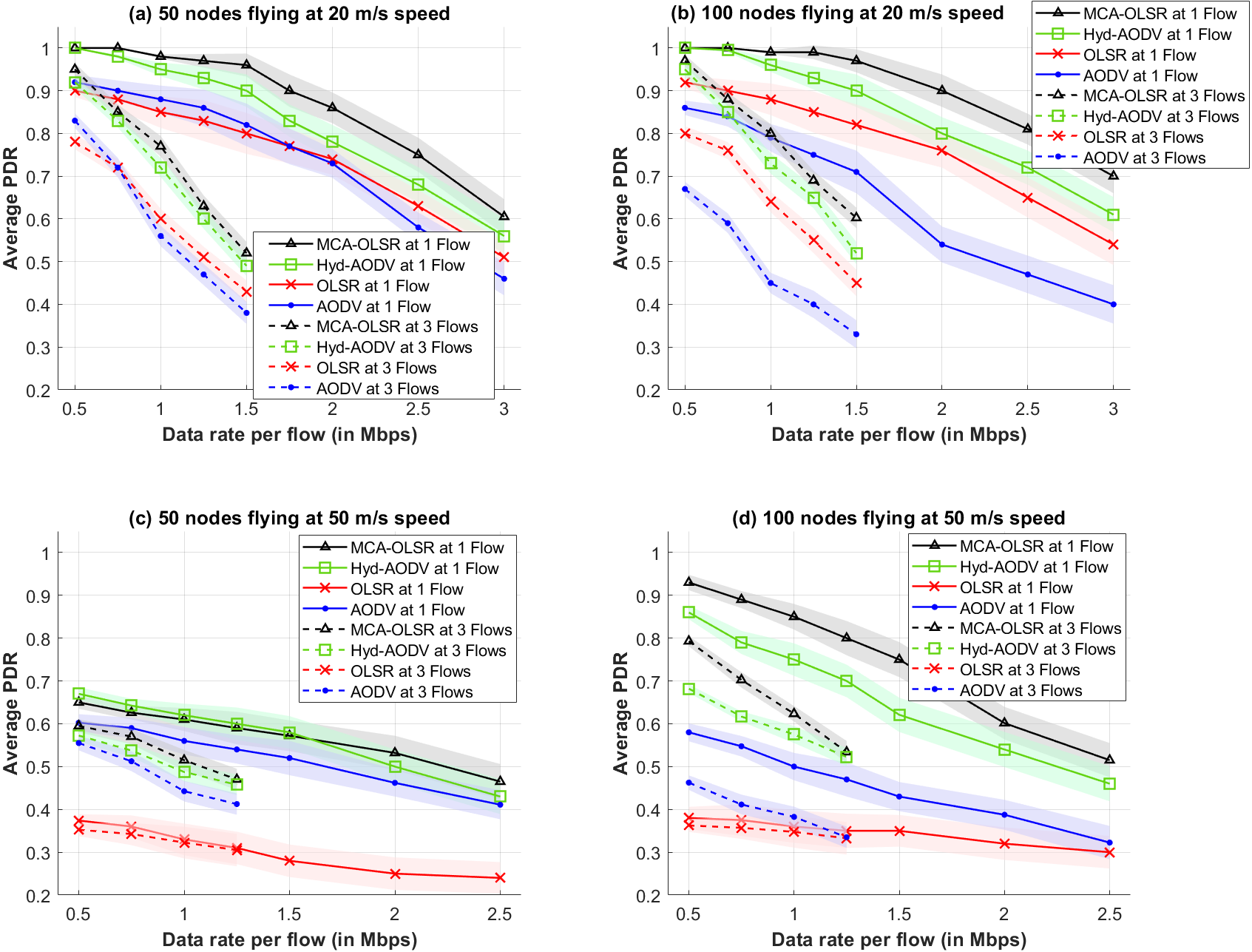}
	\caption{Average PDR for MCA-OLSR, Hyd-AODV, and standard OLSR and AODV schemes for different number of data flows and data rates, at varying node densities and speeds.}
	\label{ch5:Proactive&ReactiveComparison}
\end{figure*}

\subsubsection{Comparison of Route Control Packets}
As compared to AODV, the number of route control packets is generally lower in LEPR and Hyd-AODV schemes at different node densities, speeds and traffic loads in Fig. \ref{ch5:ReactiveSchemeCompareRemaining}. At higher node density and traffic loads, it increases significantly in AODV as compared to LEPR and Hyd-AODV schemes. When the node speed increases from 20 m/s to 50 m/s, the number of route control packets decreases significantly in AODV due to network partitioning. However, the number of route control packets changes only moderately in LEPR and Hyd-AODV schemes because majority of control packets (e.g., Notify\_Source in Hyd-AODV scheme and RSWT in LEPR scheme) are generated and forwarded along the active route(s). \\

\subsubsection{Control Overhead Comparison}
As compared to AODV, the control overhead is generally lower in LEPR and Hyd-AODV at different node densities, speeds and traffic loads in Fig. \ref{ch5:ReactiveSchemeCompareRemaining}. At higher traffic load and node density, it increases for all three schemes. However, the increase at node density of 100 nodes is much higher for the AODV scheme. At higher node speed, it decreases for all the three schemes, due to the network partitioning into multiple components. 

\subsection{Comparison with Proactive Routing Schemes}
\label{CompareWithProactiveScheme}

The choice between proactive and reactive routing protocols depends on many factors, including the network size, node mobility, and traffic characteristics. However, existing literature comparing proactive and reactive routing schemes (e.g., \cite{clausen2002comparative, leonov2018considering, alkhatieb2020performance, lau2023aqr, kim2023fanet}) offers limited and inconclusive insights. For example, \cite{clausen2002comparative} compares the performance of the standard AODV and OLSR schemes, showing the PDR and control overhead in AODV are superior to OLSR for latency-agnostic, low-throughput traffic scenarios in MANETs. But, as the traffic load increases, OLSR outperforms AODV in terms of PDR, delay and control overhead. In some FANET-based comparative studies, such as \cite{leonov2018considering, alkhatieb2020performance}, AODV outperforms OLSR in terms of PDR at all node speeds and/or traffic loads, contradicting the results in \cite{lau2023aqr}. Furthermore, in \cite{leonov2018considering}, control overhead in OLSR exceeds that in AODV when the number of traffic flows is high; this is inconsistent with the findings in \cite{clausen2002comparative}. Both AODV and OLSR schemes 
in \cite{kim2023fanet} experience lower PDR with increasing node density, yet the PDR remains consistent across various node speeds, contrary to expectations. None of the cited studies consider latency-sensitive flows with varying data rates, which can result in packet loss due to expiration or buffer overflow. 

Specific scenarios are delineated below to provide conclusive evidence of where proactive (OLSR and MCA-OLSR) and reactive (AODV and proposed Hyd-AODV) routing schemes show advantages in terms of PDR, control overhead, and route computations at different network and traffic conditions.\\ 

\begin{table*}[!t]
\centering
\caption{Comparison of Average Control (Signaling) Overhead (in MB)}
\label{ProactivevsReactiveOverheadComparison}
\renewcommand{\arraystretch}{1.5}
\resizebox{0.6\linewidth}{!}{%
\begin{tabular}{|cc|c|c|cc|cc|}
\hline
\multicolumn{2}{|c|}{\multirow{2}{*}{\textbf{Scenario}}} & \multirow{2}{*}{\textbf{OLSR}} & \multirow{2}{*}{\textbf{MCA-OLSR}} & \multicolumn{2}{c|}{\textbf{AODV}} & \multicolumn{2}{c|}{\textbf{Hyd-AODV}} \\ \cline{5-8} 
\multicolumn{2}{|c|}{} & & & \multicolumn{1}{l|}{\textbf{1 Flow}} & \multicolumn{1}{l|}{\textbf{3 Flows}} & \multicolumn{1}{l|}{\textbf{1 Flow}} & \multicolumn{1}{l|}{\textbf{3 Flows}} \\ \hline
\multicolumn{1}{|c|}{\multirow{2}{*}{\textbf{50 Nodes}}}  & \textbf{20 m/s} & 61.3 & 115.9 & \multicolumn{1}{c|}{1.9} & 6.7 & \multicolumn{1}{c|}{1.1} & 1.7 \\ \cline{2-8} 
\multicolumn{1}{|c|}{} & \textbf{50 m/s} & 5.9 & 40.9 & \multicolumn{1}{c|}{1.1} & 2.1 & \multicolumn{1}{c|}{0.7} & 0.9 \\ \hline
\multicolumn{1}{|c|}{\multirow{2}{*}{\textbf{100 Nodes}}} & \textbf{20 m/s} & 465.8 & 1,341.6 & \multicolumn{1}{c|}{33.1} & 49.6 & \multicolumn{1}{c|}{3.8} & 8.7 \\ \cline{2-8} 
\multicolumn{1}{|c|}{} & \textbf{50 m/s} & 241.7 & 588.2 & \multicolumn{1}{c|}{10.5} & 19.4 & \multicolumn{1}{c|}{2.8} & 4.9 \\ \hline
\end{tabular}}
\end{table*}

\begin{table*}[!t]
\centering
\caption{Comparison of Average Number of Routes Computed}
\label{ProactivevsReactiveComputationComparison}
\renewcommand{\arraystretch}{1.5}
\resizebox{0.6\linewidth}{!}{%
\begin{tabular}{|cc|cc|cc|cc|}
\hline
\multicolumn{2}{|c|}{\multirow{2}{*}{\textbf{Scenario}}} & \multicolumn{2}{c|}{\textbf{OLSR}} & \multicolumn{2}{c|}{\textbf{MCA-OLSR}} & \multicolumn{2}{c|}{\textbf{Hyd-AODV}} \\ \cline{3-8} 
\multicolumn{2}{|c|}{} & \multicolumn{1}{c|}{\textbf{1 Flow}} & \textbf{3 Flows} & \multicolumn{1}{c|}{\textbf{1 Flow}} & \textbf{3 Flows} & \multicolumn{1}{c|}{\textbf{1 Flow}} & \textbf{3 Flows} \\ \hline
\multicolumn{1}{|c|}{\multirow{2}{*}{\textbf{50 Nodes}}} & \textbf{20 m/s} & \multicolumn{1}{c|}{340,037} & 348,717 & \multicolumn{1}{c|}{188} & 776 & \multicolumn{1}{c|}{16} & 52 \\ \cline{2-8} 
\multicolumn{1}{|c|}{} & \textbf{50 m/s} & \multicolumn{1}{c|}{252,844} & 277,721 & \multicolumn{1}{c|}{713} & 2,606 & \multicolumn{1}{c|}{65} & 106 \\ \hline
\multicolumn{1}{|c|}{\multirow{2}{*}{\textbf{100 Nodes}}} & \textbf{20 m/s} & \multicolumn{1}{c|}{2,033,978} & 1,992,317 & \multicolumn{1}{c|}{131} & 722 & \multicolumn{1}{c|}{17} & 81 \\ \cline{2-8} 
\multicolumn{1}{|c|}{} & \textbf{50 m/s} & \multicolumn{1}{c|}{1,053,313} & 1,065,282 & \multicolumn{1}{c|}{3,274} & 8,928 & \multicolumn{1}{c|}{103} & 265 \\ \hline
\end{tabular}}
\end{table*}

\subsubsection{PDR Comparison}
Fig. \ref{ch5:Proactive&ReactiveComparison} shows the average PDR of each scheme for 1 and 3 latency-sensitive flows across varying data rates (up to the point that the network becomes congested) at different node densities and speeds. 
Although Hyd-AODV uses only a fraction of the network topology within the pipe, its PDR performance is better than standard OLSR and approaches that of MCA-OLSR.

Since fewer routes are available at lower density (50 nodes), the PDR performance of these proactive and reactive routing schemes decreases significantly when node speed increases from 20 m/s to 50 m/s (see Figs. \ref{ch5:Proactive&ReactiveComparison}(a) and \ref{ch5:Proactive&ReactiveComparison}(c)). Here the PDR of Hyd-AODV and MCA-OLSR are comparable to each other at 50 m/s, in Fig. \ref{ch5:Proactive&ReactiveComparison}(c). On the other hand, when node density is increased to 100 nodes at 50 m/s, the PDR of Hyd-AODV and MCA-OLSR increases significantly because more stable high-quality routes become available; see Fig. \ref{ch5:Proactive&ReactiveComparison}(d). 
At high node speed, the route breaks more frequently, resulting in a higher number of route discoveries in Hyd-AODV. Whereas, MCA-OLSR can select a new route in the entire network instead of being limited to the nodes in the pipe. 
Thus, it achieves higher PDR than Hyd-AODV when the node speed increases from 20 m/s to 50 m/s in Figs. \ref{ch5:Proactive&ReactiveComparison}(b) and \ref{ch5:Proactive&ReactiveComparison}(d). 

The PDR performance of the standard AODV protocol is comparable to standard OLSR on 50 nodes at 20 m/s; see Fig. \ref{ch5:Proactive&ReactiveComparison}(a). Fig. \ref{ch5:Proactive&ReactiveComparison}(b) shows the PDR performance for 100 nodes at 20 m/s. Due to higher node density (100 nodes) and low node speed (20 m/s), more stable routes are available between the source-destination pairs, improving the PDR performance of standard OLSR compared to the lower node density.
However, the PDR performance of standard AODV decreases due to the increased control overhead for more nodes; see Fig. \ref{ch5:ReactiveSchemeCompareRemaining}.

Since standard OLSR selects a shortest hop route and uses the edge nodes, it experiences frequent link breaks at high node speed (50 m/s), giving short-lived routes. In contrast, standard AODV experiences fewer route breaks because it may not select the shortest hop route available in the network{\footnote{The source node selects the shortest hop route to the destination node only from among the received RREP packets, as discussed in Section \ref{BasicAODV}.}.
This property enables AODV to achieve a higher PDR than standard OLSR at higher speeds; see Figs. \ref{ch5:Proactive&ReactiveComparison}(c) and \ref{ch5:Proactive&ReactiveComparison}(d).\\

\subsubsection{Performance for 10 Sensor Data Flows} Table \ref{10Flows} shows the PDR performance comparison for 10 sensor data flows at two different data rates for both node densities and speeds. Here, the PDR trends are similar to those observed for one and three flows discussed previously: the Hyd-AODV scheme significantly outperforms standard OLSR and is comparable to MCA-OLSR  except at high node density and speed.\\

\subsubsection{Control Overhead Comparison}
The control overhead includes the Hello and TC packets in OLSR, and RREQ, RREP and RERR packets in AODV, besides the Notify\_Source packets in Hyd-AODV and MCA-OLSR schemes. The control overhead in both AODV and Hyd-AODV schemes is significantly lower than the standard OLSR and MCA-OLSR schemes for both node densities and speeds in Table \ref{ProactivevsReactiveOverheadComparison}. While the control overhead for proactive routing schemes is independent of the number of traffic flows, it increases with traffic load for both AODV and Hyd-AODV schemes. Due to network partitioning into multiple components at 50 m/s, the control overhead for all the schemes is lower as compared to 20 m/s.\\

\subsubsection{Average Number of Route Computations}
Table \ref{ProactivevsReactiveComputationComparison} shows the average number of routes computed by both proactive and reactive routing schemes. Since standard AODV does not compute routes at the source node (instead, it simply selects the route conveyed by the RREP packet(s)), it is not included in the table. Both MCA-OLSR and Hyd-AODV recompute the routes only when a new, better quality route becomes available, while the standard OLSR protocol recomputes the route whenever a new control packet is received. Since Hyd-AODV monitors the routes only within the pipe, which is a fraction of the network topology, the total number of routes computed ($T$) is significantly lower than for MCA-OLSR and standard OLSR. This results in lower computational complexity of $O(T {V_p}^2 (E_p+V_p))$ (see Section \ref{CompComplexity}). Here, the number of nodes $V_p$ and edges $E_p$ within the pipe are significantly less than the total number of nodes $V$ and edges $E$ in the network.\

\section{Discussion}
\label{Keypoints}
The proposed Hyd-AODV scheme achieves higher instantaneous as well as average PDR than the AODV, OLSR and LEPR schemes. It selects stable and less congested routes, preemptively switches to a better route when the current route quality degrades, and uses a proactive queue management policy. Despite using only limited network topology information, Hyd-AODV often provides PDR that is comparable or only slightly below that of MCA-OLSR.

At higher node speed, the network experiences frequent topology changes and network partitioning, resulting in situations where no route may be available for a flow, decreasing the PDR of all the evaluated routing schemes. Hyd-AODV uses the $RLT$ metric in route selection and switching, while AODV does not track $RLT$ and OLSR often selects edge nodes in its shortest $HC$ route. Thus, Hyd-AODV achieves higher PDR than AODV and OLSR at higher node speeds. Since LEPR triggers a new route discovery after its cached route(s), found during the previous route discovery, become obsolete, it often incurs delays.

A higher node density increases the number of available stable routes, improving the PDR of Hyd-AODV, MCA-OLSR and LEPR, especially at higher node speeds. Since the network topology stays more connected at lower speeds, the PDR gain from higher density is slight. 

Hyd-AODV has fewer route discoveries than  AODV and LEPR because it searches for an alternate route only within the pipe before initiating a new route discovery. The route discoveries increase with traffic load in all three reactive routing schemes (Hyd-AODV, AODV and LEPR). 
The number of route control packets transmitted by Hyd-AODV and LEPR are much lower than in standard AODV. 
However, because of the additional fields introduced for both Hyd-AODV and LEPR, their total control (signaling) overhead (in MB) are only somewhat lower than AODV at lower node density. 
At higher density, the control overhead in AODV is much higher than for Hyd-AODV and LEPR. However, the control overhead in all three reactive routing schemes is much lower than in the proactive routing schemes, OLSR and MCA-OLSR.

Since Hyd-AODV computes a route only using the pipe information, its route computation overhead and the resulting complexity are much lower than for OLSR and MCA-OLSR.

The proposed scheme uses the following thresholds: $\delta$ for route longevity, $\epsilon_1$ and $\epsilon_2$ for route delay, weights $w_1$ and $w_2$ in route cost computation, and $W$ for pipe width. 
A node pair computes the $LLT$ value when they exchange trajectory information via the Hello packet; therefore $\delta \geq$ Hello interval. Increasing $\delta$ enforces a stricter route longevity constraint, which can reduce the available routes, and increase the number of route switches and discoveries at higher node speeds.
$\epsilon_1$ and $\epsilon_2$ are used for route selection and switching, respectively, with $\epsilon_1 > \epsilon_2$. Increasing these values prioritizes nodes with lower $PST$, which reduce latency. However, at high traffic loads, fewer routes may meet this constraint, leading to more route switches and discoveries.
Under constraint $w_1 + w_2 = 1$, increasing $w_1$ favors shorter-hop routes, while increasing $w_2$ prioritizes routes with lower interference.
Increasing pipe width $W$ provides more high-quality routes, which may improve throughput and reduce total route discoveries. However, it also increases the overhead incurred in collecting and sending neighborhood information to the source node. In a multi-flow network topology, a larger $W$ may cause interference with neighboring flows, which would reduce the flow throughput and increase the number of route discoveries.

The energy consumed is not considered since (per experimental studies \cite{abeywickrama2018comprehensive, zeng2017energy}) the energy for communication is several orders of magnitude smaller than for keeping the UAV aloft. However, UAVs can fail due to battery depletion or malfunction. In the event of such node failures, the availability of enough alternate routes (see Table \ref{TradeOffOverheadVsHQROutes}) means that the proposed scheme can easily switch to a new high-quality route.

\section{Conclusion}
\label{Conclusion}
A mobility and congestion-aware hybrid reactive routing protocol (Hyd-AODV) was proposed for decentralized UAV networks. 
The proposed protocol searches high-quality route(s) on-demand (using a multi-metric route selection), monitors the dynamic region (called the pipe) around the selected route(s), and proactively switches to an alternative route (within the pipe) before the quality of the current route degrades below a threshold. An AoI-aware queue management scheme was also incorporated to prioritize the transmission of packets. 
The impact of pipe width on the ability to find alternate high-quality routes within the pipe, as well as on route selection, switching, queue management, and the required overhead, was studied.

The Hyd-AODV scheme achieves superior performance, in terms of throughput, number of route discoveries, overhead, and flow interruptions, compared to the standard AODV and LEPR routing schemes. Despite using only limited network topology information, the flow throughput performance of  Hyd-AODV scheme was superior to standard OLSR and close to MCA-OLSR on a range of network and traffic settings, while incurring significantly lower control (signaling) overhead and computational complexity. Compared to existing literature, a far more detailed evaluation of the relative performance of reactive and proactive routing schemes was provided across different network and traffic settings. 

In the future, the routing scheme will be extended to study the impact of varying channel characteristics and external interference on its performance. Other future extensions of the proposed Hyd-AODV routing scheme could include integration with different UAV swarm mobility models (such as the pheromone-based mobility model). In addition, a topology control mechanism could be designed to form alternate high-quality routes within the pipe to improve flow reliability and network resiliency.

\vspace{\baselineskip}
\noindent\textbf{Acknowledgment of Support and Disclaimer:\ } 
The authors acknowledge the U.S. government’s support in the publication of this paper. This material is based on work funded by the U.S.\ AFRL (Air Force Research Laboratory), under  Contract  No. FA8750-20-1-1005 and FA8750-23-C-0521. Any opinions,  findings and conclusions or recommendations expressed in this material are those of the author(s) and do not necessarily reflect the views of AFRL or the US government.

\ifCLASSOPTIONcaptionsoff
    \newpage
\fi

\bibliographystyle{IEEEtran}

\end{document}